\documentclass{revtex4}

\usepackage{graphicx}
\def\beq{\begin{equation}}
\def\eeq{\end{equation}}
\def\bea{\begin{eqnarray}}
\def\eea{\end{eqnarray}}
\def\ee{e^+e^-}
\def\chp#1{\widetilde\chi^{~+}_{#1}}
\def\chm#1{\widetilde\chi^{~-}_{#1}}
\def\chpm#1{\widetilde\chi^{~\pm}_{#1}}
\def\chmp#1{\widetilde\chi^{~\mp}_{#1}}
\def\chz#1{\widetilde\chi^{~0}_{#1}}

\def\sell{\widetilde e_L}
\def\selr{\widetilde e_R}

\def\smupr{\widetilde \mu^{~+}_R}
\def\smumr{\widetilde \mu^{~-}_R}
\def\smul{\widetilde \mu_L}
\def\smur{\widetilde \mu_R}

\def\stau#1{\widetilde \tau_{#1}}

\def\snue{\widetilde \nu_e}
\def\snum{\widetilde \nu_\mu}
\def\snut{\widetilde \nu_\tau}

\begin{document}
% You should use BibTeX and revtex.bst for references
\bibliographystyle{revtex}

% number  and your conference paper identification number on the
% title page in preprint mode. Multiple \preprint commands are allowed.
\preprint{ }

\title{Experimental Approaches at Linear Colliders}
%\title[]{}

\author{Marco Battaglia}
\email[]{Marco.Battaglia@cern.ch}
\affiliation{CERN, CH-1211 Geneva 23, Switzerland}
\author{Ian Hinchliffe}
\email[]{I_Hinchliffe@lbl.gov}
\affiliation{LBNL, Berkeley, CA, USA}
\author{John Jaros}
\email[]{john@slac.stanford.edu}
\affiliation{SLAC, Menlo Park CA, USA}
\author{James Wells}
\email[]{jwells@physics.ucdavis.edu}
\affiliation{UC Davis, Davis CA, USA}

\date{\today}

\begin{abstract}
\vspace*{0.25cm}

\end{abstract}
% \pacs{}

%\maketitle must follow title, authors, abstract and \pacs
\maketitle

\section{Introduction}

Precision measurements have played a vital role in our understanding
of elementary particle physics. Experiments performed using $e^+e^-$
collisions have contributed an essential part. Recently, the precision
measurements at LEP and SLC have probed the standard model at the
quantum level  and severely
constrained the mass of the Higgs boson~\cite{lepewwg:2001}.
Coupled with the limits on the Higgs mass from direct searches
\cite{lephwg:2001}, this enables the mass to be constrained to be
in the range 115-205~GeV.  Developments in accelerator
R\&D have matured to the point where one could contemplate
construction of a linear collider with initial energy in the 500~GeV range
and a credible upgrade path to $\sim 1$ TeV. Now is therefore the
correct time to critically evaluate the case for such a facility.

The Working Group E3, {\it Experimental Approaches at Linear Colliders}, was 
encouraged to make this evaluation.
The group was charged with examining critically the physics case for a
Linear Collider (LC) of energy of order 1~TeV as well as the cases for
higher energy machines, assessing the performance requirements and exploring the 
viability of several special options. In addition it was asked to identify the 
critical areas where R\&D is required (the complete text of the charge can be found 
in the Appendix).
In order to address this, the group was organized into subgroups, each
of which was given a specific task. Three main groups were assigned to
the TeV-class Machines, Multi-TeV Machines and Detector Issues. The central
activity of our working group was the exploration of TeV class machines, 
since they are being considered as the next major initiative in high energy
physics. We have considered the physics potential of these machines, the special 
options that could be added to the collider after its initial running, and
addressed a number of important questions. Several physics 
scenarios were suggested in order to benchmark the physics reach of the
linear collider and persons were appointed to maintain contacts with the
relevant activities in the various Physics Working Groups.  Special
options considered were
precision electroweak studies that could be done by running the
collider at and near the $Z$ pole (so called Giga~Z running);  collisions
involving $\gamma\gamma$, $e^-e^-$, or $e \gamma$ interactions;
and positron beam polarization. The following questions were posed in
order to focus the discussions: 
(1)  In view of the fact that the luminosity is a function of energy,
what are the trade-offs involved in selecting the energy.
(2) What is the argument for proceeding with the construction of a
Linear collider as soon as possible rather than waiting for data
from LHC? (3) In the context of a definite physics scenario, what is a
realistic run plan? {\i.e.} How much luminosity at each energy? (4)  What
should be the initial energy of a linear collider and  to what energy 
should that machine extended? 

While the reports from these various activities should be consulted for 
details~\cite{schumm,Barklow:2001mm,dq4-report,grannis-paper,Erler:2001su,
Velasco:2001vg}, the group summary will cover
the most important activities and conclusions of the group. While we
have tried to reflect the consensus of the working group, the conclusions 
expressed in this report are the responsibility of the authors.

Exploiting the full potential of a Linear Collider requires a powerful
detector. The conceptual detector designs that have emerged from the
{\sc Tesla}~\cite{tesla-tdr},NLC~\cite{nlc-resource} and JLC~\cite{Abe:2001gc} 
are aggressive evolutions from the LEP and SLC detectors. These designs
are driven by the requirements of precision measurements of the
properties of a Higgs boson, reconstruction of final states involving
$W$ and $Z$ bosons, and the precise mass measurement of new
particles. These issues occupied many of the groups discussions~\cite{schumm} 
with a view to identifying the critical areas where R\&D is needed.

As many important issues such as the experimental environment and 
the performance trade-offs involve machine physics, there were several joint 
sessions with the M3 group. In particular, we wanted to obtain a clear understanding
of the accelerator R\&D issues that need to be resolved before the
proposed machine could be constructed. 

This
report covers the activities of the group on TeV-class machines in
some detail and provides the overall group summary by drawing on the
work of the other subgroups. Section~II is devoted to the energy frontier and 
the expected physics
thresholds based on our current knowledge of the Standard Model (SM).
 Section~III covers the important detector issues and the 
special options listed
above. Section~IV is devoted to issues associated with
the run scenarios and Section V with the more esoteric physics
possibilities.
Section~VI is devoted to the ultimate energy of the machine and the
possibilities for achieving it. Finally, Section~VII has our
conclusions.

\section{New Physics Thresholds at the LC}

Historically, elementary particle physics has  gained important new
insights into  the fundamentals of the structure of nature at  high energies  
by precisely measuring the interactions between known
particles and by discovering new particles.  
As has been documented  in many published reviews, and as will
be further illustrated in these proceedings, a linear collider is
an exceptional machine to make precision measurements provided it has
sufficient energy so that new thresholds can be reached.

Foremost, there are good indications that a light Higgs boson can be studied
at such a collider. There is compelling evidence  that
the Higgs boson will be accessible to a machine with energy in the 500~GeV range. 
Firstly, precision electroweak measurements
such as the $Z$-pole measurements at LEP and SLC
of total $Z$-width, forward-backward asymmetry, left-right asymmetry,
etc.\ have put serious constraints on any theoretical description
of electroweak symmetry breaking (EWSB). Many proposed models of new physics
are excluded  because they
have not passed the test of making correct predictions for these
observables.  Other models  have been given more credibility by
being 
compatible with the precise data.  The $W$ boson mass measurements at
LEP \cite{lepw:2001},
CDF\cite{cdfw} and D0 \cite{d0w} and the
top quark mass measurement at CDF \cite{cdftop} and D0 \cite{d0top}
are also important additions to the  constraints on theories.

\begin{figure}[h!]
\includegraphics[scale=0.4]{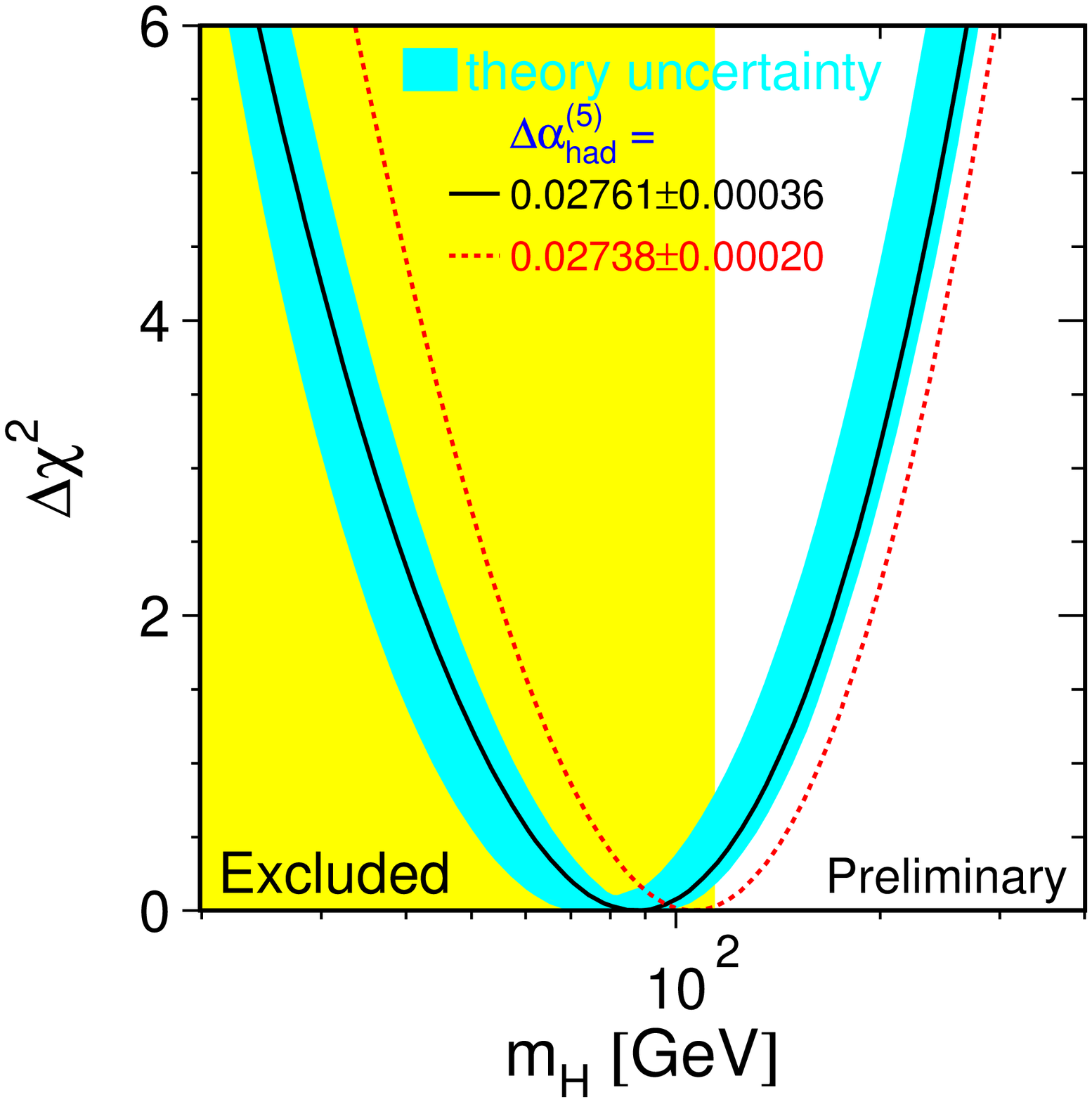} 
\caption{The $\chi^2$ of the fit to the electro-weak observables
  measured 
at LEP, 
SLC and Tevatron as a function of the Higgs boson mass. The preferred value is 
$M_H = 88^{+53}_{-35}$~GeV and the one-sided 95\%~C.L. upper limit is
196~GeV (from~\cite{lepewwg:2001})}
\label{fig:e3001_fig1}
\end{figure}

The standard model with a single Higgs boson  works very well as a description of
the data, provided the Higgs boson mass is below about 200~GeV 
(see Figure~\ref{fig:e3001_fig1}).  
Since Higgs bosons can be produced in association
with the $Z$-boson at $e^+e^-$ colliders, the predicted mass is well
within the kinematically  accessibile range of a 500~GeV linear collider.
Precise studies of this new threshold (Higgs boson)
associated with EWSB will be available  at a linear collider.  

Models can be constructed  that include
additional states and are capable of conspiring with a heavy Higgs 
boson to mimic the effects of a light Higgs boson in the precision
electroweak data.  
An exhaustive review was carried out~\cite{Peskin:2001rw}, 
and it is expected that most such models 
require exotic states (scalar pseudo-Nambu-Goldstone bosons,
$Z'$, etc.) which are detectable at a 500~GeV linear collider.  
In some cases, there would be
no directly observable state at this energy.  However, more 
precise measurements of the $W$ mass and implementation of the
GigaZ option would  see deviations from the standard model, and would
be able to yield information about the nature of the new physics.  
In all cases, the linear collider adds crucial information
to previous data and to the data that the LHC will obtain.

The unification of the three gauge couplings works to impressive
precision in supersymmetric theories, giving strong support for
perturbative grand unified theories incorporating supersymmetry.
  In these theories, an upper bound on the lightest
Higgs mass can be obtained~\cite{Kane:1993kq}, of  205~GeV~\cite{Espinosa:1998re}.  
This bound
is almost the same as the 95\% upper limit from precision electroweak
data, which of course is derived independently.  

The standard model   must be embedded  into a more
fundamental theory valid at higher energies if an explanation of its
many parameters is to be obtained.  
In such models. the $W$ mass is very sensitive to  new high
scales, and therefore  it is expected that the dynamics associated with electroweak
symmetry breaking be not too far above 1~TeV.  Otherwise,
large numbers anchored to a large high-scale must conspire to cancel out
and  give a small number of order $\sim m_W$.  
  Differences arise  due to  uncertainty in what physics is to be associated with 
this new more fundamental scale $\Lambda$ and can affect the precise
value. It could be the 
scale below which the Higgs
boson is a composite particle, and above which no fundamental
scalars exist.  Or it could be the new scale  where
superpartners exist, completing the states needed for a supersymmetric
description of nature that protects scalars from receiving dangerous
quadratic divergences.  

The supersymmetric interpretation is the
simplest perturbative explanation for low-scale electroweak
symmetry breaking, and its minimal model predicts that the lightest
Higgs boson must be below about 130~GeV, independent of the 
precision electroweak data or the gauge coupling unification argument
given above.  Furthermore, since supersymmetry is a highly predictive
and highly developed theory, many studies have been made to quantify
the naturalness requirements that the weak scale places on the
superpartner masses.  Upper limits disagree in their precise values, but
the general consensus is that some superpartners should be 
accessible at a 500~GeV linear collider (see Chapter 4, Sec.\ 2 of
Linear Collider Resource Book~\cite{nlc-resource}).

One new threshold is known to be accessible a  linear collider.
The top quark mass is
35~times larger than any other known quark, and about 100 times larger than
any known lepton. It is the only fermion that has an ${\cal O}(1)$ Yukawa
coupling to the standard model Higgs boson. 
 Its strong Yukawa coupling makes it
unique among fermions.  Studying all observables associated with this
quark may yield critical insights into a more fundamental
theory of its mass generation, and electroweak symmetry breaking in general.

Finally, we remark that there is the logical possibility that no Higgs
boson exists, fundamental or composite, 
and that strong dynamics sets in at the TeV~scale. It is not clear how the
precision electroweak data would be satisfied in such a theory, but it
would probably need to involve other light states~\cite{Peskin:2001rw}
(see above).  In the unlikely event that such light states 
are not present, a linear collider still has 
capabilities in detecting subtle behavior in
longitudinal vector-boson
scattering, e.g.\ $e^+e^-\to \nu\bar\nu W^+W^-$ and $W$ pair
production from $e^+e^-\to W^+W^-$ 
Signal significance at a 500~GeV collider could be comparable 
 to that of the LHC (see sect.\ 3.1 of the
Linear Collider Resource Book~\cite{nlc-resource}).

\section{LC Experimental Issues and Special Options}

The linear collider must be able to  provide critical data which will enable 
the  answers to today's major questions on the mechanisms responsible for the breaking 
of some of the symmetries in nature, on the origin of mass and on the nature of the new 
physics to be obtained. This will require flexibility of 
operation, high luminosity and a carefully designed interface with the experimental 
apparatus. A significant flexibility will be necessary for adapting the modes of 
operation to the nature of the new physics signals to be investigated. This implies the 
ability to vary the center-of-mass energy over a wide range, to perform closely 
spaced energy scans of resonances and to collide different kinds of beams in different 
polarization states. High luminosity will allow the accuracy achievable in determining 
the properties of the phenomena under study to be at the  level required to identify 
their nature and to complement the LHC measurements.
Finally the interface between the accelerator and the detector requires special care for 
achieving the needed accuracy in the reconstruction of the events, despite the 
backgrounds originating in the delivery and interaction of the intense beams and also 
the final focus stabilization system.

\subsection{Machine Constraints and Performance}

A linear collider operating at energies around 500~GeV has the sensitivity to study the 
new phenomena as discussed in the previous section. This center-of-mass energy is 
reachable by accelerator technologies that are ready for large scale 
implementation. Therefore 500~GeV can be assumed as the primary initial energy of a 
TeV-class LC. At this energy the LC will be sensitive to most of the Higgs boson 
properties and should be able to detect and untangle signals of new physics, either 
directly or indirectly.

\begin{figure}[t]
\includegraphics[scale=0.45]{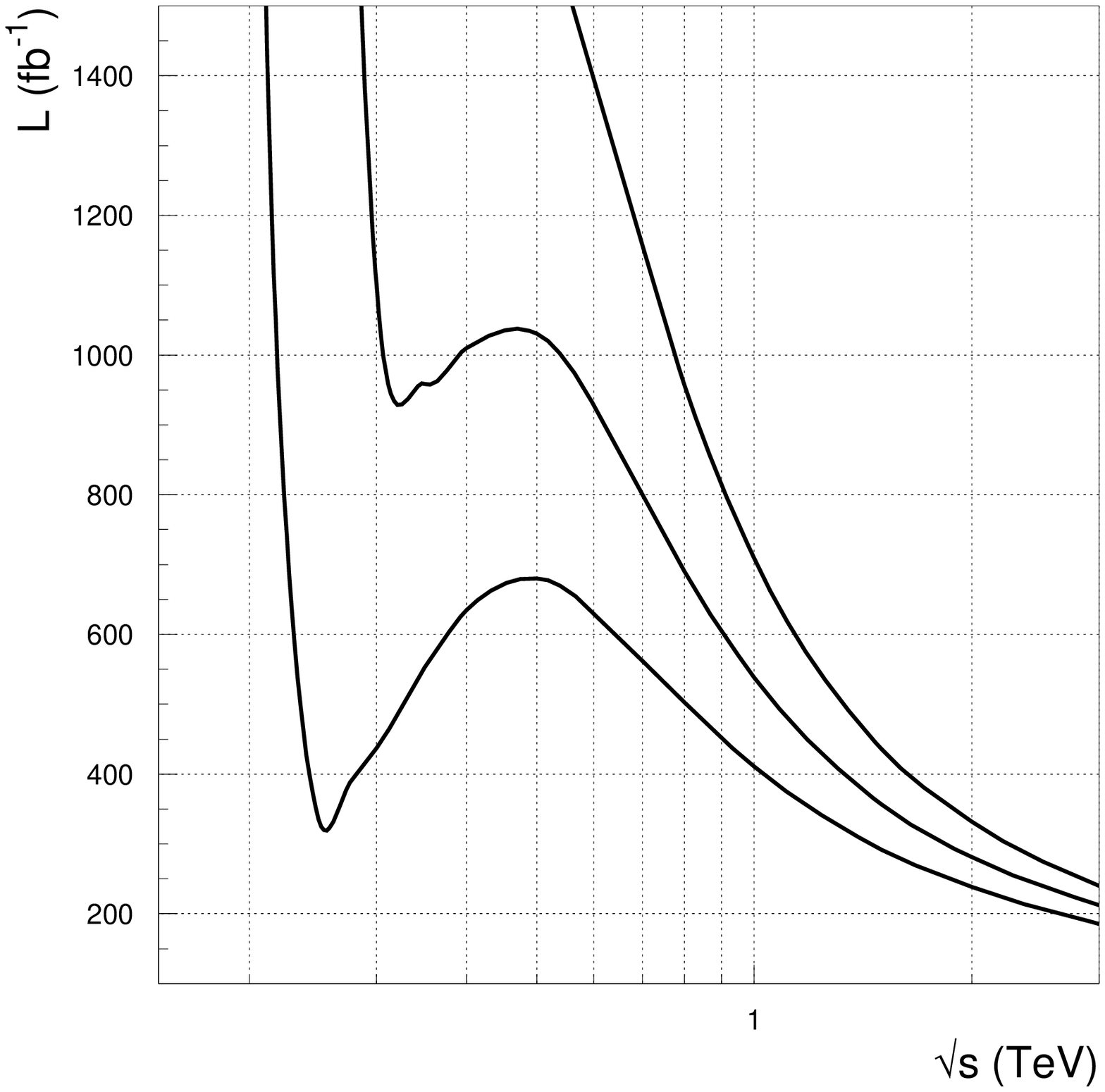} 
\caption{The scaling of the integrated luminosity, necessary to produce 10$^5$ Higgs 
bosons in $e^+e^-$ collisions, with the center-of-mass energy $\sqrt{s}$ for three 
values of the Higgs mass: 120~GeV, 180~GeV and 240~GeV.}
\label{fig:e3001_fig2}
\end{figure}

While the precise center-of-mass energy requirement of further stages of LC operation 
will be refined by the details of the phenomenology as it will emerge from 
the {\sc Tevatron} and LHC data, indications already exist. 
At present, the highest energy known to correspond to the excitation of a particle 
threshold to be studied at the LC is the $t \bar t$ threshold around 350~GeV. 
This motivates the $\sqrt{s}$=350~GeV energy for a phase of the LC operation. 
The scan of the top production threshold requires about 100~fb$^{-1}$ of integrated 
luminosity to achieve an statistical accuracy on the top quark mass of
$\simeq$~50~MeV~\cite{martinez}, comparable to the systematic uncertainty of
$\simeq$~50~MeV~\cite{Hoang:2000yr} from theoretical sources.
In the context of the SM, where the Higgs boson is expected to be light, as indicated 
by the electro-weak data and of Supersymmetry a similar energy also corresponds to the 
peak of the associated $H^0Z^0$ production rate. However, a detailed study of the Higgs 
boson requires more luminosity. Figure~\ref{fig:e3001_fig2} shows the integrated 
luminosity needed to produce 10$^5$ Higgs bosons through either of the Higgs-strahlung or
$WW$ fusion processes. Such a sample guarantees a statistical accuracy of the order of 
few \% in the determination of the principal Higgs couplings to force and matter 
particles. 
Therefore a luminosity of at least to $10^{34}$cm$^{-2}$s$^{-1}$ needs to be achieved 
in this phase. 

Apart from the exploration of the top threshold and of possible other thresholds that 
may appear up to 500~GeV, energy and integrated luminosity can be traded to obtain a 
given sample of signal events. 
At energies up to about 3~TeV the s-channel is the dominant process for 
production of pairs of fermions and gauge bosons. Since the corresponding cross-sections
decrease as $1/s$, operating at higher energies requires  even larger luminosities. This
is, for example the case for the important $H^0Z^0$ associate production process, as 
highlighted in Figure~\ref{fig:e3001_fig1} which shows the luminosity required to 
produce a constant sample of $H^0$ bosons increases for energies above about 300~GeV. 
Eventually the $WW$ fusion process becomes dominant around 0.8-1.0~TeV.

\subsection{Physics Constraints and Detector Concept}

The defining features of $e^+e^-$ collisions, i.e. the known energy, $\sqrt{s}$, the 
known identity of the initial state partons, and the opportunity to control the 
helicities of the initial state partons, apply to .5-1 TeV collisions essentially as 
they did at LEP and SLC energies.

To be sure, high energy collisions at a linear collider involve extremely intense 
beam-beam 
interactions, which produce beamstrahlung and secondary beamstrahlung-electron/positron 
interactions at levels far exceeding those at lower energies.  This radiative process, 
like the initial state radiation inevitable in $e^+e^-$ interactions, causes a radiative
tail in 
the collision energy.  But, since the effects are rather comparable to those occurring 
naturally, 67\% of the collisions are still within 1\% of the nominal EM energy, and 90\% 
are within 5\%.  For all intents and purposes, the center of mass energy is still 
well-defined at a TeV-class LC

\begin{figure}
\includegraphics[scale=0.75]{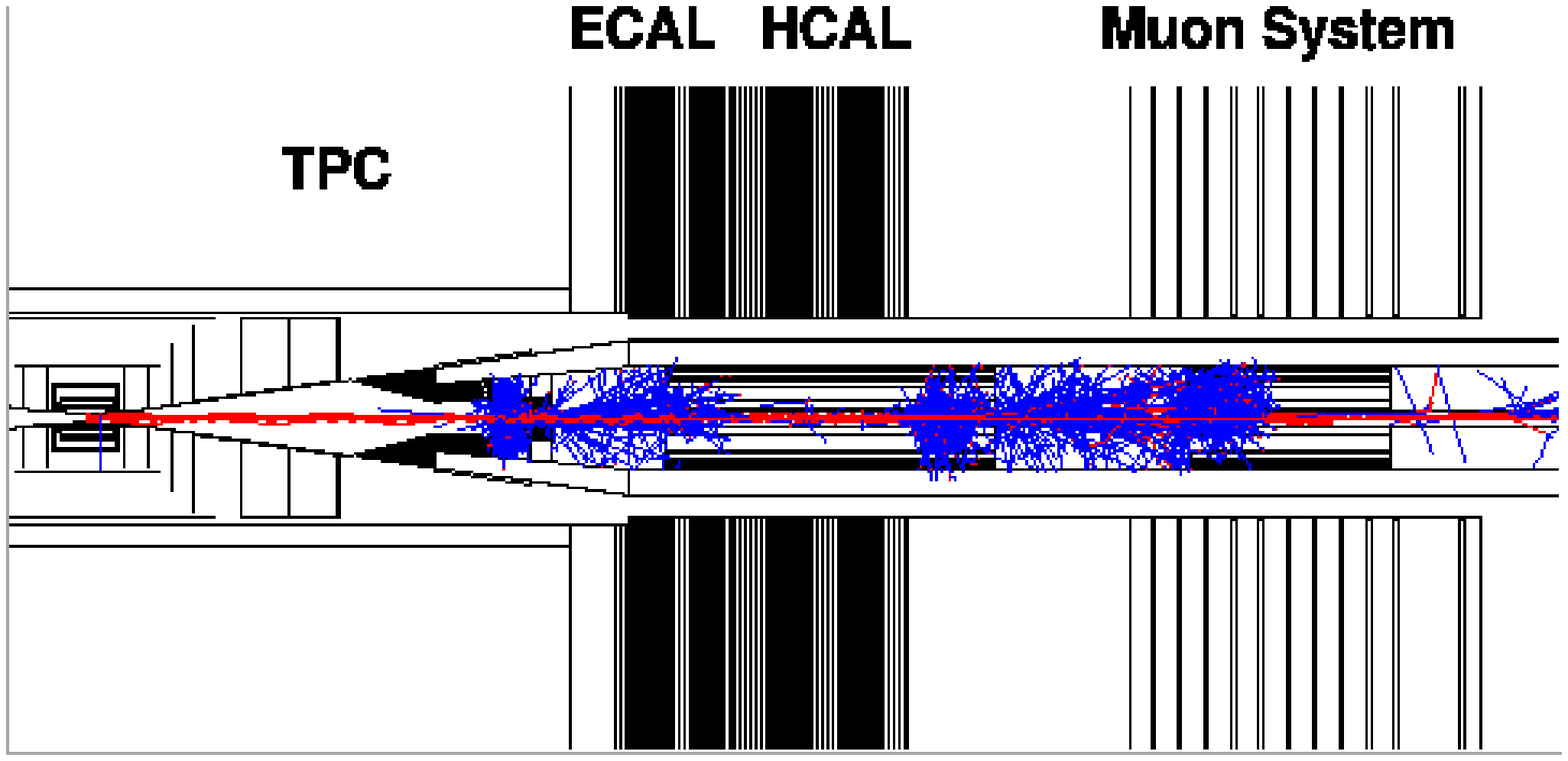} 
\caption{The LC interaction region with the $e^+e^-$ pair background confined by the 
solenoidal magnetic field and shielded by a system of forward masks.}
\label{fig:e3001_fig3}
\end{figure}

The physics production cross-sections are electroweak at the LC, so the raw event rate and 
the resultant radiation field are low, and there is probably no need for a trigger. 
The beamstrahlung involving the interaction of an electron with the opposing electron
bunch causes copious production of $e^+e^-$ pairs, but present detector designs, which 
incorporate multi-Tesla solenoidal magnetic fields, capture the pairs close to the beam line,
and divert them away from the Interaction Point (see Figure~\ref{fig:e3001_fig3}).  
This strategy is so successful that the first vertex layer can be placed within nearly a
centimeter of the beamline.  The LC environment is benign compared to that of hadron colliders.
Detector designs can be optimized for physics performance, not radiation hardness or 
high-rate capability. A broad range of detector technologies can be considered.

Reaching the full potential of a TeV class linear collider will require new detector 
technologies to be developed and existing technologies to be refined.   
The designs are driven by the 
requirements of precision Higgs physics, precise measurements of the masses of 
supersymmetric particles, and the clean identification of $W$, $Z$, Higgs, and top by 
reconstructing multi-jet final states.

\subsubsection{Vertex Detection at the LC}

The Vertex Detector (VXD) will provide fermion flavor tagging, which is
needed to understand electroweak symmetry breaking and the mechanism of
mass generation.
Precision measurements of the Higgs boson couplings to different fermion species 
are essential to test that the  Higgs mechanism is indeed responsible for the 
generation of mass. Accurate determinations of the decay 
rates to $b\bar{b}$, $c\bar{c}$, $\tau^{+}\tau^{-}$, and gluon pairs will determine
whether the Higgs coupling is indeed proportional to mass.
Determining these branching fractions with high precision may also distinguish 
Standard Model behavior from that expected in its extensions, such as supersymmetry.
Since the rates for the Higgs decay modes into lighter fermions or into gluon pairs are expected 
to be only about 10\% or less of that for the dominant $H^0 \rightarrow b \bar b$ process, the 
extraction and measurement of these signals requires a strong suppression of the 
$b \bar b$ contribution while maintaining  good efficiency 
(see Figure~\ref{fig:e3001_fig5}). Distinguishing charm and
light flavors from bottom by identifying the decay vertex and determining its 
multiplicity and reconstructed mass~\cite{hansen}, puts a premium on improvements to 
impact parameter resolution, especially for low momentum tracks.

In order to fulfill these challenging requirements, the LC VXD aims to improve 
the current state of the art several fold to achieve impact parameter resolution 
of 3-5~$\mu$m for high momentum tracks, and to reduce the multiple
scattering contribution to the momentum resolution 
to $\sim$5~$\mu$m/$p$~(GeV/c). 
%Obtaining high precision requires pure and efficient isolation of Higgs decay modes which 
%are relatively rare, especially charm and gluon decays, and that in turn requires truly 
%superb vertex detection.  New designs boost the already impressive impact parameter 
%resolution achieved by SLD's CCD vertex detector by a factor of three, giving resolutions 
%of about 3~$\mu$m for high momentum tracks 
Two environmental constraints are significant~\cite{Battaglia:2000kd}.  
The accelerator-induced background at the radius of 
the innermost sensitive layer causes a hit 
density of the order of 0.2-1.0~hits~mm$^{-2}$, larger than those expected at the LHC. 
This, and the track density in highly collimated jets,  
require detectors with small pixel cells and relatively fast read-out or time stamping 
capabilities. 
The flux of neutrons incident on the innermost layers of the vertex detector is of the order of 
$10^9$cm$^{-2}$year$^{-1}$, which is large enough to be of concern for CCD sensors.  

\begin{figure}
\includegraphics[scale=0.5]{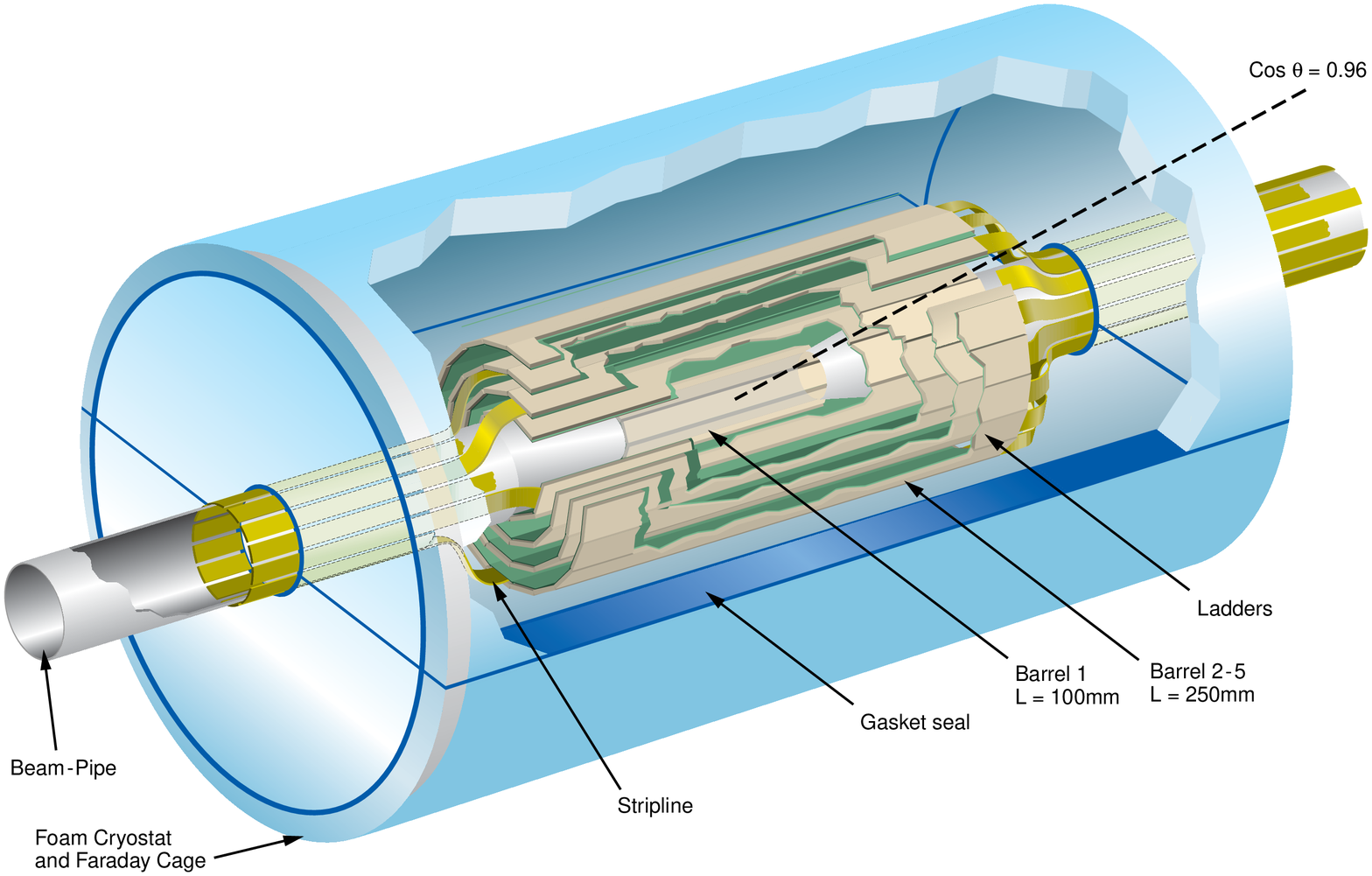} 
\caption{Isometric view of the Vertex Detector surrounding the interaction region. The 
concentric layers of high resolution Si pixel detectors ensure the discrimination of 
charged particles produced in the decay of short-lived heavy hadrons from those 
originating at the point of beam interaction (from~\cite{tesla-tdr}).}
\label{fig:e3001_fig4}
\end{figure}

Several VXD designs have been proposed, relying on different sensor technologies 
(see Figure~\ref{fig:e3001_fig4}). All these designs rely on pixel devices, and all
try to minimize the detector thickness to improve the impact parameter resolution for
low momentum tracks.
A substantial R\&D effort is underway already for CCD 
detectors~\cite{Damerell:2001kh,burrows-ccd}, aiming at radiation 
hardening~\cite{Brau:2000pt} and boosting readout speeds by developing new 
architectures.  Other pixel detector technologies are also being evaluated, including 
monolithic CMOS detectors~\cite{Claus:2001bq,deptuch}, in which readout electronics is 
incorporated 
directly on high purity silicon; and bump-bonded pixel sensors~\cite{Battaglia:2001nd}.  

\subsubsection{Central and Forward Tracking}

\begin{figure}
\begin{tabular}{l r}
\includegraphics[scale=0.4]{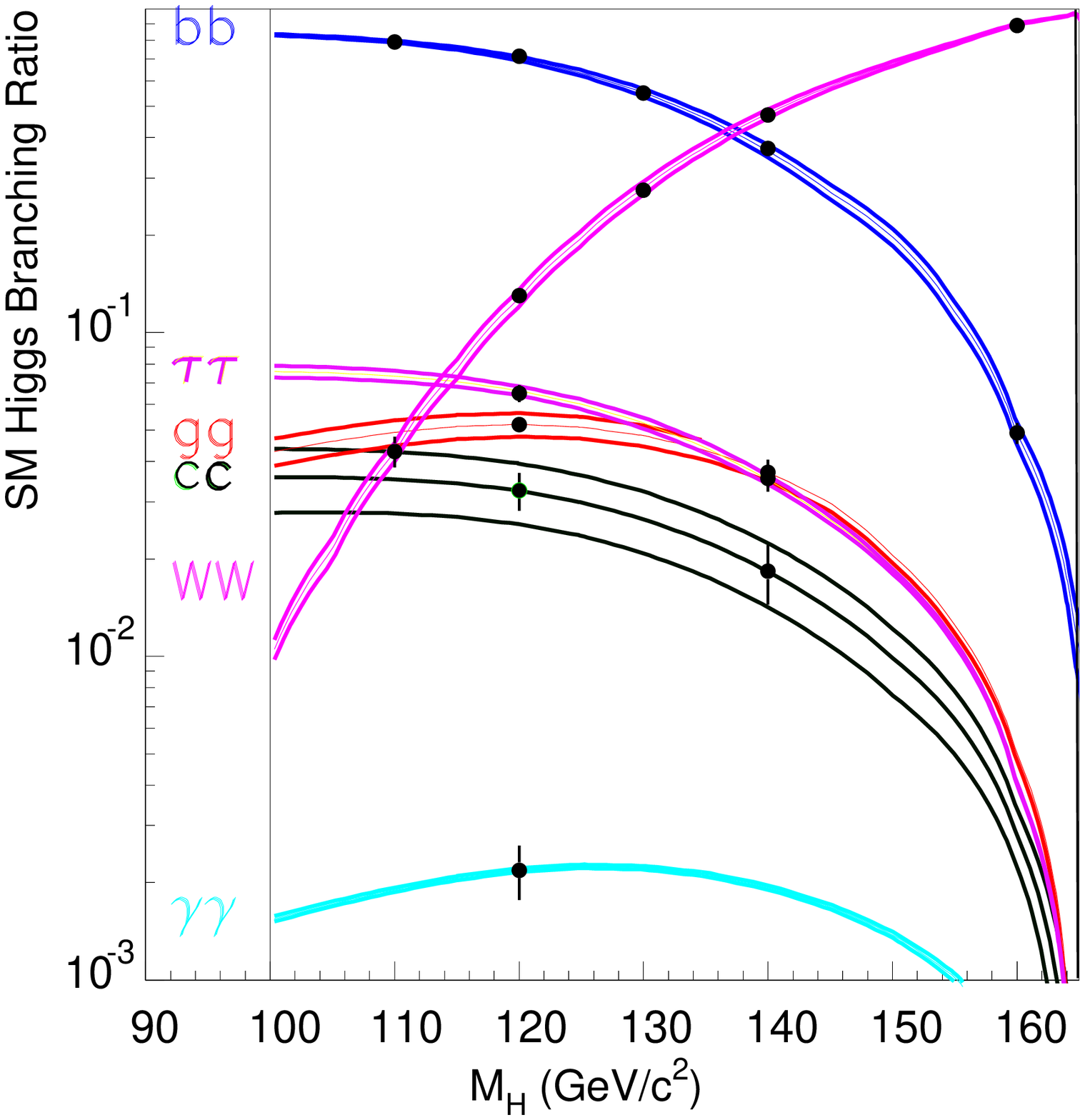} &
\includegraphics[scale=0.4]{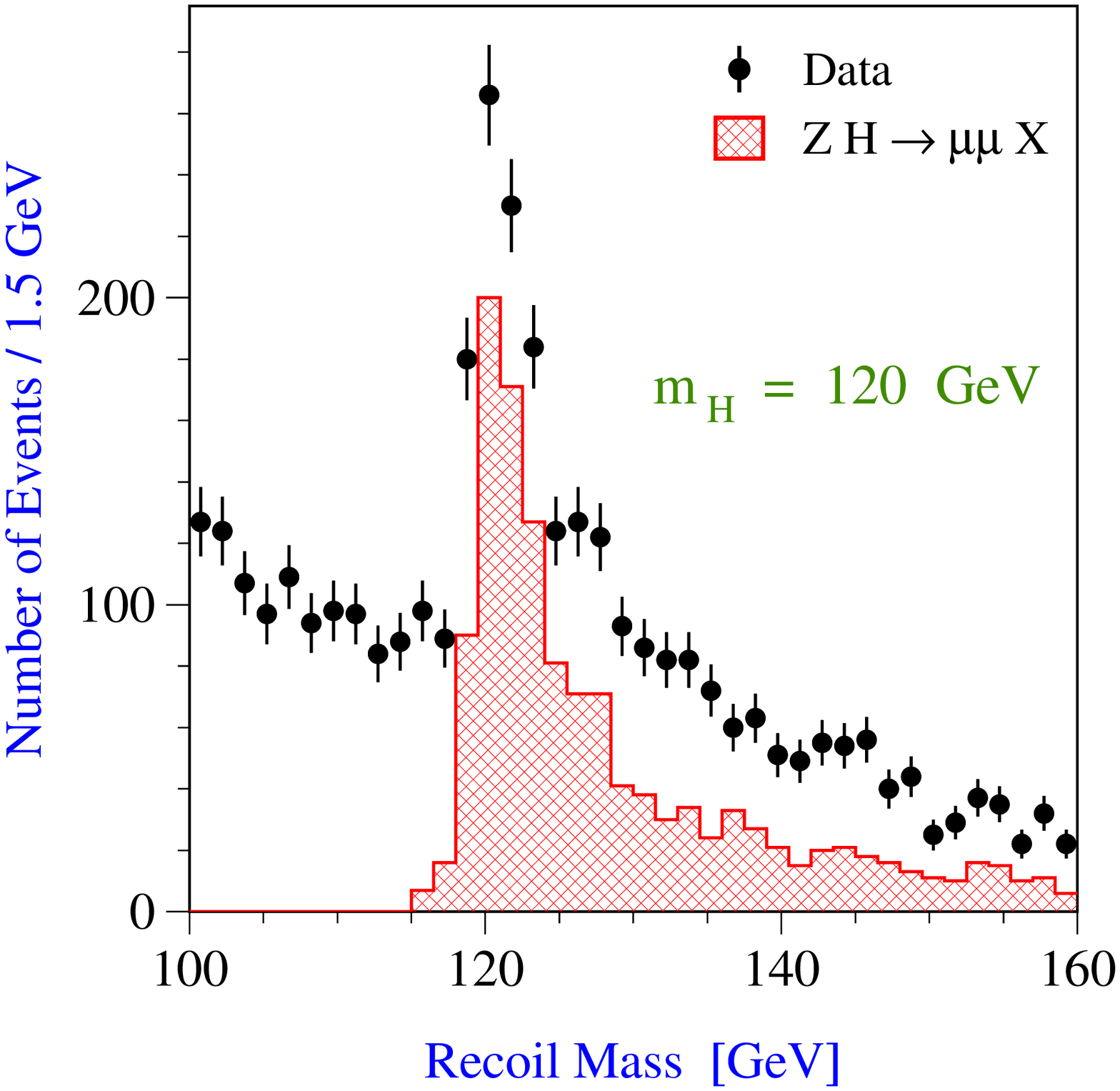} \\
\end{tabular}
\caption{Left: Higgs decay branching fractions predicted in SM as a function of the 
Higgs boson mass. The lines represent the expected range due to uncertainties in the 
input parameters and the points with error bars represent the estimated statistical 
accuracies obtainable at a LC with 500~fb$^{-1}$ (from~\cite{Battaglia:2000jb}). 
Right: reconstructed recoil mass spectrum showing the mass peak due to a 120~GeV Higgs 
boson produced in the $e^+e^- \rightarrow ZH$ process (from~\cite{tesla-tdr}).}
\label{fig:e3001_fig5}
\end{figure}

\begin{figure}
\includegraphics[scale=0.45]{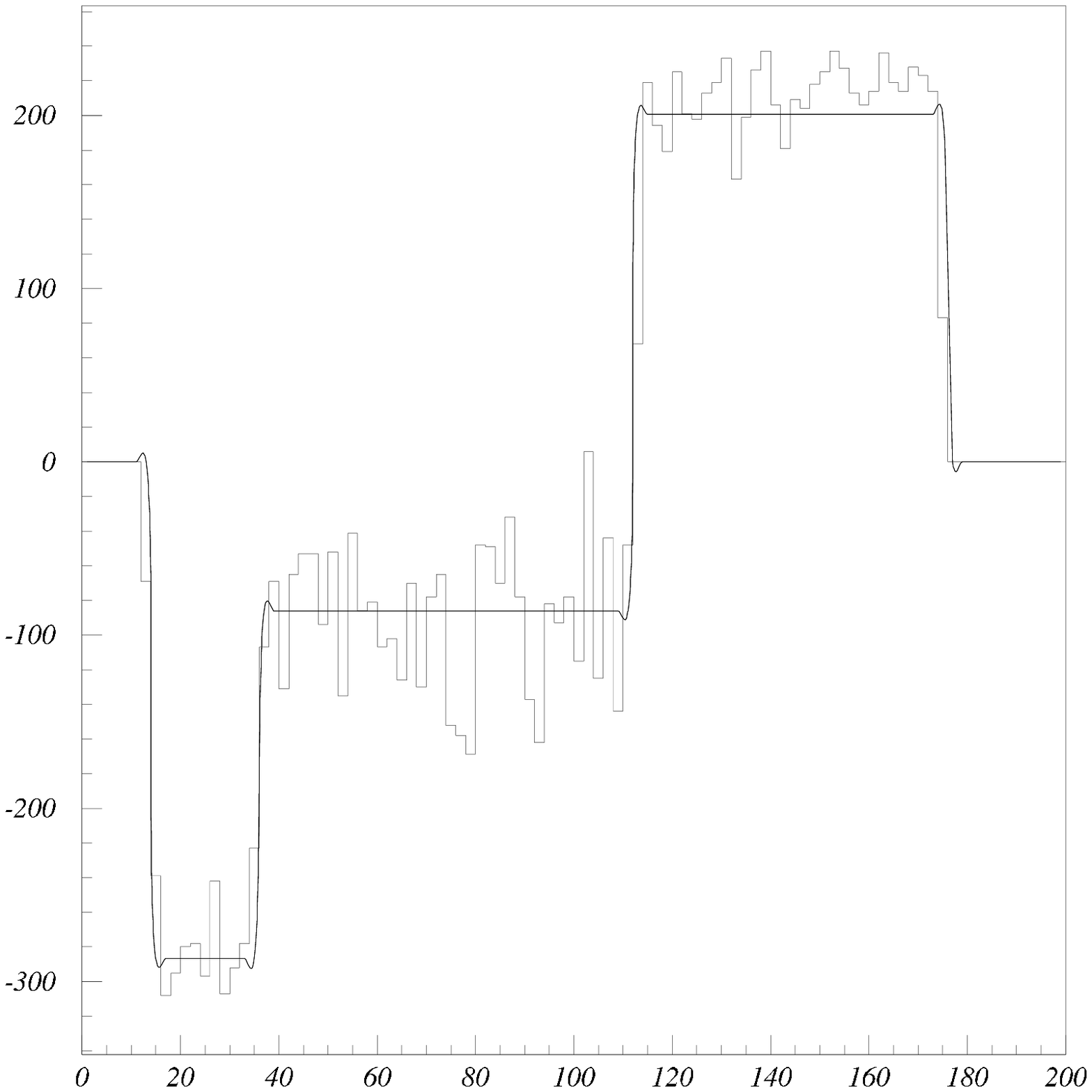}
\caption{Example of determinations of masses of sleptons from the measured momentum 
distribution of muons detected in events produced with different beam polarization 
states (from~\cite{Dima:2001jr}).}
\label{fig:e3001_fig7}
\end{figure}

Higgs physics, and the study of supersymmetric particles, both demand extremely high momentum 
resolution.  In Higgstrahlung ($e^+e^- \rightarrow H^0Z^0$), the model-independent 
identification of the Higgs boson depends on the accuracy with which the recoil mass to the 
$Z^0$ can be reconstructed from the dilepton decay $Z \rightarrow \ell^+ \ell^-$ 
(see Figure~\ref{fig:e3001_fig5}).  
Th accuracy in the recoil mass is fundamentally limited by beamstrahlung and beam energy 
spread, which broadens the spike in the center of mass energy to the 0.2\% level.  
To saturate that limit, the momentum resolution for charged particles should be pushed 
towards $\delta p/p^2\simeq10^{-5}$GeV$^{-1}$, an order of magnitude
beyond what has been achieved at LEP.  
Such momentum resolution would also sharpen the distinctive box spectra 
expected from two body SUSY decays, thus improving the accuracy on the mass measurement 
of supersymmetric particle partners (see Figure~\ref{fig:e3001_fig7}).

Additional requirements on the tracking include pattern recognition capability in 
the presence of backgrounds, adequate track separation resolution to resolve all the 
tracks in the dense core of a high energy jet and full solid angle coverage.
Several technologies for central tracking are presently being considered to meet these 
challenging requirements.  The study in ref.~\cite{Abe:2001gc} has 
concentrated on a traditional drift chamber design.  Issues such as electrostatic stability, 
gravitational sag, and cell design have already been addressed with detector R\&D.  Pattern 
recognition capability in the presence of backgrounds must still be evaluated.  
Ref.~\cite{tesla-tdr} has advanced the idea of a large volume Time Projection Chamber 
(TPC), 
extending radially between about 30 and 170~cm, with about 120 radial measurements.  The 
high granularity of the sensitive cells in the gas volume provides excellent pattern 
recognition capabilities, even in presence of backgrounds.  
Present designs of trackers incorporating a TPC offer good momentum resolution, 
$\delta p/p^2 = 5 \times 10^{-5}$~GeV$^{-1}$, if critical systematics and calibration issues 
can be adequately addressed~\cite{Behnke:2001gb}. 
The inherent $dE/dx$ information available in a TPC also gives some particle ID capability. 
R\&D is underway on new ways to detect the drifted electrons~\cite{Behnke:2001pd}, including 
GEM~\cite{Sauli:1997qp} or Micromega~\cite{Giomataris:1996fq} detectors replacing the 
traditional MWPCs; on optimal gas mixtures~\cite{Gruwe:1999bd}; on integrated readout 
electronics; and on thinning the TPC endplate~\cite{tpc-report}.  
A different approach, advocated 
in~\cite{nlc-resource} is to rely on a 
multi-layered silicon tracker, based on either silicon microstrip or silicon drift sensors.  
Such a device has excellent momentum resolution ($\delta p/p^2 = 2 \times 10^{-5}$~GeV$^{-1}$),
but reduced stand-alone pattern recognition capability.  The robustness and the efficiency of
the combined vertex detector + silicon tracker pattern recognition against the expected 
backgrounds and efficient enough in reconstructing $K^0_s$ and $\Lambda^0$ have yet to 
be demonstrated.  
R\&D is proceeding on several items: long and short shaping time options for 
the silicon detectors; the effect of high magnetic fields on the detectors; and power cycling
to match the very low LC duty cycle and detector alignment.

Many tracking issues remain, some of which are under active investigation.  
The track-pair resolution achieved for each technology is being tested in energetic and highly 
collimated jets. The usefulness of recording timing information, in order to 
associate tracks with a  particular bunch crossing is being investigated. The importance of thinning 
TPC endcaps, and designs for doing so, are under study, as are designs for chambers which would
reside behind the endcap to boost resolution for forward going tracks. 
A high precision tracking
layer located  between the vertex detector and the central tracker (an "intermediate tracker") has 
been proposed to boost momentum resolution. It needs further study. Forward
tracking systems, which must have excellent resolution in collinearity angle for the measurement
of the differential luminosity spectrum, need further development.

\subsubsection{Calorimetry}

\begin{figure}
\begin{tabular}{l r}
\includegraphics[scale=0.45]{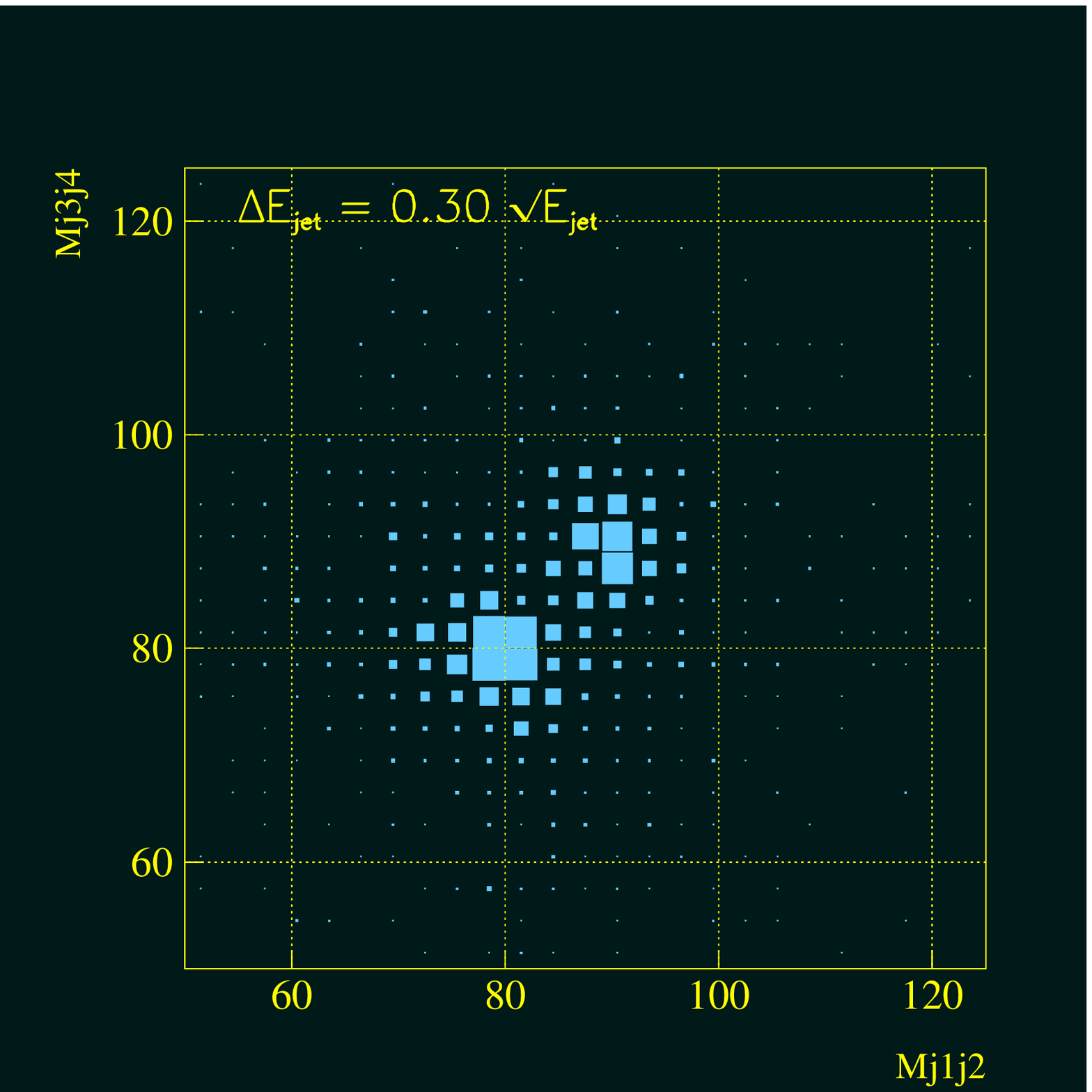} &
\includegraphics[scale=0.45]{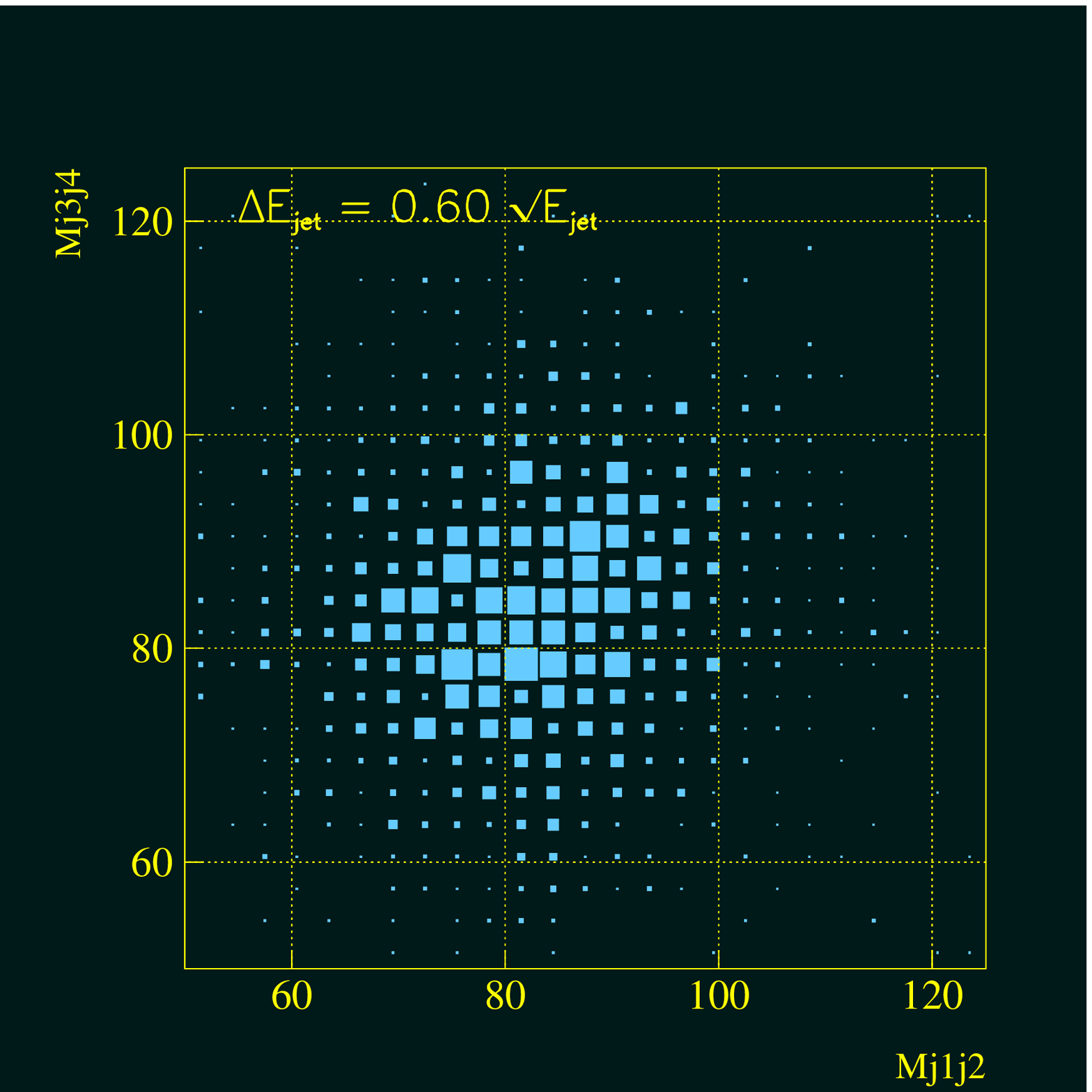} \\
\end{tabular}
\caption{An exemplification of the impact of jet energy resolution on the separation of 
$WW$ from $ZZ$ hadronic decays: the masses of each di-jet pairs are shown for 
$60\%/\sqrt{E}$ (right) and $30\%/\sqrt{E}$ (left) (from~\cite{videau:2001xx}).}
\label{fig:e3001_fig9}
\end{figure}

\begin{figure}
\includegraphics[scale=0.45]{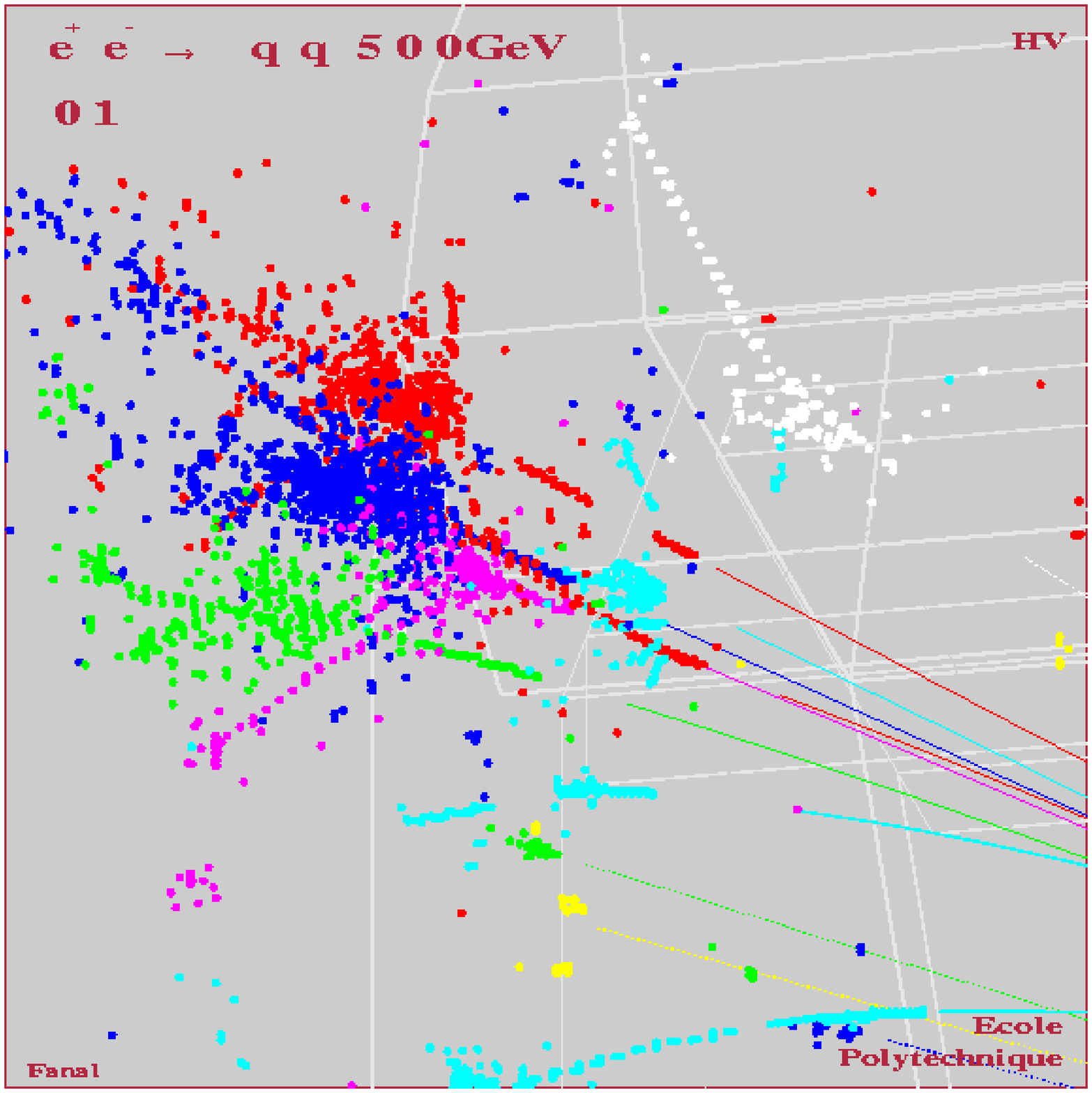}
\caption{The response of a tungsten digital hadronic calorimeter to Particles from an hadronic jet (from~\cite{videau:2001xx}).}
\label{fig:e3001_fig10}
\end{figure}

The goal of the calorimetry is to unambiguously identify $W$, $Z$, Higgs bosons, and top 
quarks on the basis of the invariant mass of multi-jet combinations 
(see Figure~\ref{fig:e3001_fig9}).  To do so will require significant advances in jet 
energy resolution and calorimetric pattern recognition. Several ideas are being explored
for improving jet energy resolution~\cite{videau:2001xx,morgunov}. The most topical, and 
promising, is that of energy-flow calorimetry, in which one measures the electromagnetic 
showers in such fine-grained 3-D cells, that the energy deposits arising from 
charged tracks penetrating the calorimeter are separable from the electromagnetic energy 
clusters (see Figure~\ref{fig:e3001_fig10}). 
If this is done, the charged energy can be measured with high precision in the 
central tracker, the electromagnetic energy with moderate precision in the electro-magnetic 
calorimeter (ECAL),  and the remaining neutral hadronic energy in the hadronic calorimeter 
(HCAL).  The resolution on the sum improves significantly in this approach.  
In the ideal case, where there is no confusion in assigning the energy to charged and 
neutral components, the jet energy resolution is $18\%/\sqrt{E}$, in contrast with that 
expected with a calorimeter-only resolution of $64\%/\sqrt{E}$, even if the calorimeter 
is fully compensating. Studies indicate that improvements in the HCAL resolution are 
very important to improve the overall jet energy resolution. The concept of a "digital 
calorimeter", in which the fine-grained cells of the hadronic calorimeter simply record
if they have been hit, has been proposed~\cite{tesla-tdr,videau:2001xx} and looks 
promising.

The study of energy flow calorimetry is dependent on full and accurate simulations of 
shower development, and well-developed pattern recognition and clustering software. 
These  are still in their infancy, leaving much to be done toward 
understanding what realistic performance gains might be and how best to optimize the 
detector.  Although results are still preliminary and no full agreement between different 
algorithms has been reached, a jet resolution of $26\%/\sqrt{E}$ appears achievable with 
$30\%/\sqrt{E}$ resolution in the HCAL and $15\%/\sqrt{E}$ resolution in the ECAL.
Such a resolution is 
adequate for distinguishing $W$ from $Z$ bosons or $Z$ from low mass $H$ bosons, which
is not possible with jet energy resolutions in today's calorimeters.
Ongoing R\&D looks into optimizing the longitudinal segmentation vs cost and 
studying the effect of the dead channel fraction on the ultimate performance and cost of 
the device.

Energy-flow calorimetry depends on isolating charged and neutral particles, which is 
best 
done in a dense detector medium where shower spread is minimized.  Silicon/tungsten is 
favored, but its high cost demands understanding tradeoffs in pixel size, 
longitudinal segmentation, and detector quality~\cite{Brient:2001am,videau:2001xx}.  
It will be necessary to develop 
distributed electronics that can be integrated with the silicon detectors, and to solve 
the mechanical problems associated with constructing such a detector.
More traditional calorimeters are also being considered for use at the LC such as 
$5 \times 5$ cm$^2$ scintillator tiles interspersed with
Fe, readout via wavelength shifting fibers by Avalanche photodiode pixel arrays. A 
prototype calorimeter is under construction.  Other studies~\cite{Abe:2001gc,matsunaga} 
have focused
on a nearly compensating lead/scintillator design for the ECAL,  which is segmented 
longitudinally into three parts. The device has already been beam-tested.

\subsubsection{Other Systems}

The LC detector will certainly require muon identification.  Preliminary designs for 
the muon system have emerged, but work is needed to select an optimal detector 
technology. 
The LC physics may, particularly for GigaZ studies,  benefit from dedicated particle 
identification.
For example, in $B$ mixing or asymmetry measurements at the $Z$, identified kaons 
can be used to tag the flavor of  the parent $B$ hadron.
The overall value of particle ID in the high energy LC environment is still uncertain, 
but needs further investigation~\cite{soffer}.

A linear collider detector also comprises subsystems uncommon to present $e^+e^-$ 
storage rings 
detectors, some of which will need significant R\&D.  Extreme forward calorimetry, 
designed to veto the two-photon backgrounds to selectron searches, must isolate high 
energy electrons from large, but more diffuse electromagnetic backgrounds. 
A pair monitor 
will provide the instantaneous luminosity if it can separate signal from background  
close to the beam line.  The beam energy will likely be determined by use of a dedicated 
spectrometer; getting the desired level of precision will require R\&D~\cite{torrence}. 
Beam polarization must be monitored, presumably by Compton backscattering polarized 
laser light.

\subsection{LC Special Options}

\begin{figure}[t]
\begin{tabular}{c c}
\includegraphics[scale=1.35]{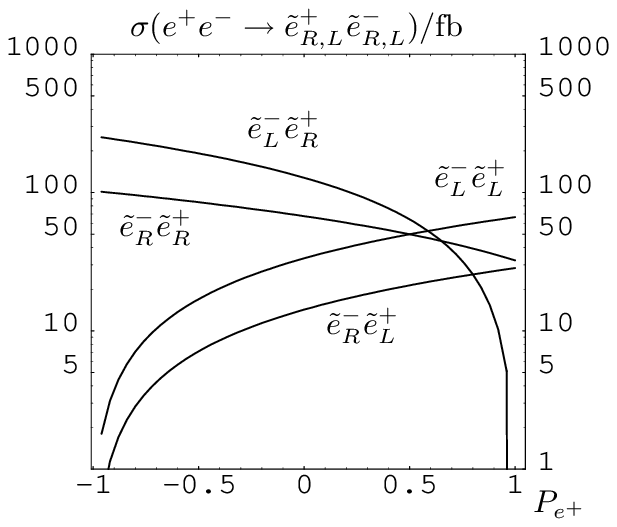} &
\includegraphics[scale=0.4]{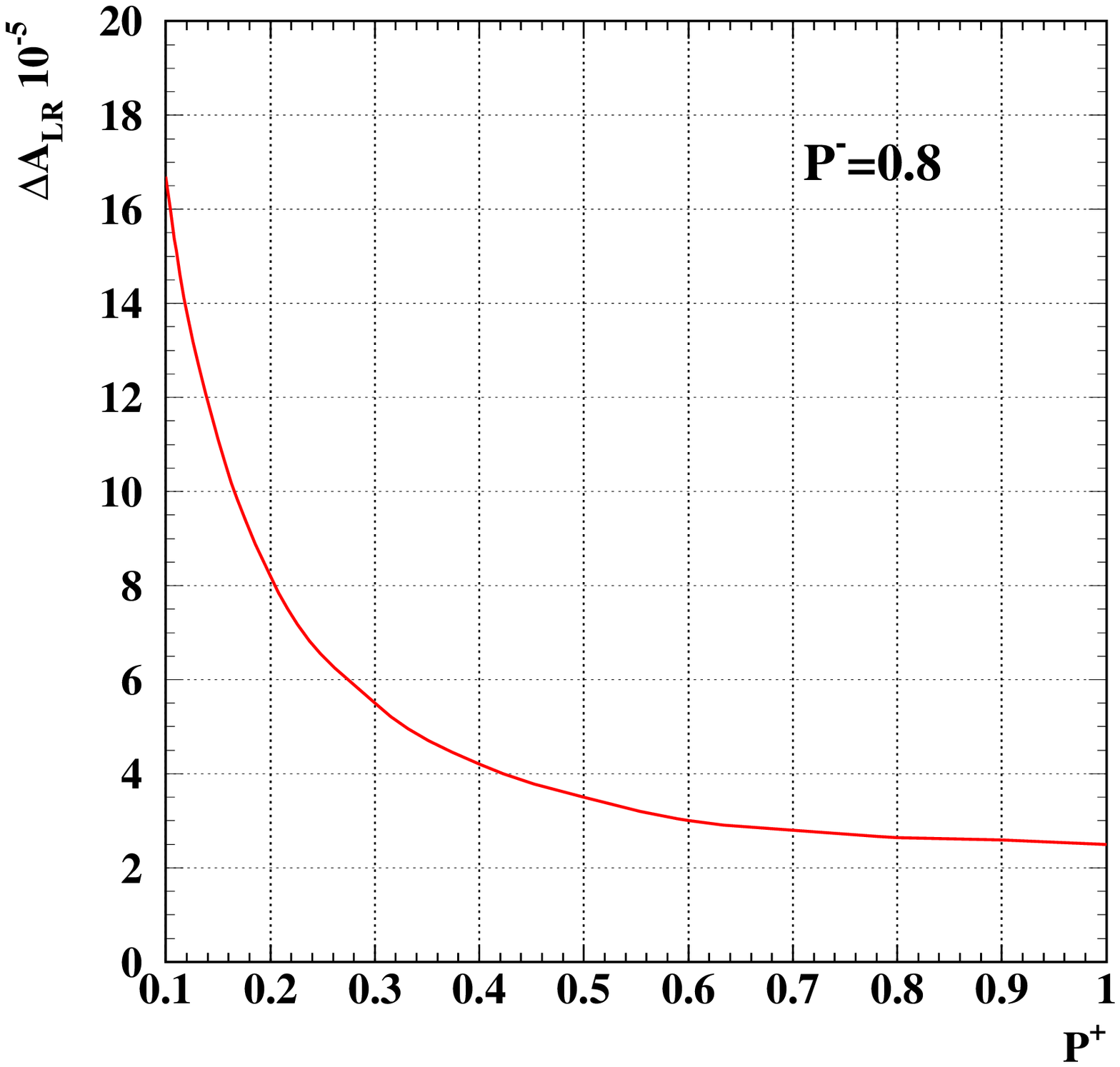} \\
\end{tabular} 
\caption{Two examples of the advantages of $e^-$ and $e^+$ polarisation in studying 
new physics and performing precision tests of the SM. Left: cross section for the 
production of the supersymmetric electron partners as a function of the positron beam
polarisation ${\cal{P}}_{e^+}$ (from~\cite{moortgat:2001xx}). 
Right: statistical uncertainties for left-right 
asymmetry $A_{LR}$ in the $e^+e^- \rightarrow Z^0 \rightarrow \ell^+\ell^-$ reaction, 
measured operating the LC at the $Z^0$ pole energy, as a function of the positron 
beam polarisation (from~\cite{Erler:2001su}.}
\label{fig:e3001_fig11}
\end{figure}

The capability of the LC to interact elementary particles with well defined energy and 
quantum numbers is further enhanced by the availability of polarised 
beams~\cite{moortgat:2001xx}. Colliding 
electrons in a given polarisation state allows tests of the helicity structure of the 
reactions originated by the particle collision. While the SM processes are mostly 
initiated by electrons polarised in the same directions, such as $e^+_Le^+_R$, new 
physics, as SUSY, may be manifested in the interactions of $e^+_Le^+_L$ and 
$e^+_Re^+_R$. This allows to efficiently separate these physics processes from those due 
to standard interactions. The use of beam polarisation carries three major benefits: 
i)~enhance the production rate of signal processes, ii)~improve the signal-to background 
ratio by reducing the rate of SM backgrounds, iii)~provide an analysers of the 
quantum numbers of the particles produced (see Figure~\ref{fig:e3001_fig11}). 
In order to fully exploit these benefits, 
it is important to ensure the implications of the LC operation with polarised beams.
The feasibility of operating a linear collider with a large degree of polarisation of 
the electron beam has been demonstrated at the {\sc Slc}, where 
${\cal{P}}_{e^-}$=0.76\% has been routinely achieved. Measurement of the polarisation 
within 0.5\% has also been demonstrated. While this is presently considered as a possible
upgrade at a later phase of LC operation, there is now ongoing R\&D addressing the 
issues related to obtaining polarised positron beams of sufficient intensity. 
For the LC the 
simultaneous polarisation of the electrons and positrons has important benefits. 
This enhances the analysing power since the effective polarisation 
${\cal{P}}_{eff} = {\cal{P}}_{e^-}-{\cal{P}}_{e^+}/(1-{\cal{P}}_{e^-}{\cal{P}}_{e^+})$ 
exceeds the degree of polarisation achievable for each single 
beam~(see Figure~\ref{fig:e3001_fig11}). Further the 
exact degree of the beam polarisation during the interactions can be extracted directly 
from the data recorded. This will largely reduce the systematic uncertainties, otherwise 
expected to dominate the overall achievable measurement accuracy.
But a second aspect has to be taken into account which concerns the impact of operating 
with polarised
beams on the achieveable luminosity. The anticipated decrease in luminosity has to be 
factored in when accounting for the enhancement of signal cross sections and highlights 
the main benefit of positron polarisation from its use as analyser.

The operation of the LC as a $\gamma \gamma$ collider may significantly extend its 
physics reach~\cite{Velasco:2001vg}. 
The principle of colliding photon beams, generated by scattering 
intense laser beams on the electron bunches, was proposed since long. Recently the 
technical implications and the engineering solutions for a realistic design of the 
$\gamma\gamma$ interaction region and the intense laser system have been investigated 
in great details. This has provided a proof of principles for this option. The point of 
strength of the photon collider largely rely on its ability to concentrate the full 
collision energy in the creation of a single spin-0 Higgs boson. On the contrary in 
$e^+e^-$ collisions these bosons are pair produced, either with a $Z^0$ or with 
another Higgs particle. By gaining access to the effective $H \gamma \gamma$ coupling, 
that involves loop of virtual particles, much can be learned on new particles even if 
their masses happen to be too large to be directly produced. The single Higgs production 
also ensures that the sensitivity to the heaviest of these particles can be pushed to 
higher masses compared to $e^+e^-$ collisions.

For some specialised physics cases there are advantages in operating the LC colliding 
electrons on electrons, for shorter periods. A good example is offered by the superior 
capability of a $e^-e^-$ collider in determining the $\tilde{e_R}$ mass due to the
significantly sharper onset of the $\tilde{e_R}\tilde{e_R}$ threshold as discussed below.

\section{Running Scenarios at the LC}

The control of the beam energy at a linear collider will allow 
complicated new physics to be disentangled and precisely measured, one threshold at a 
time, if necessary.  But skeptics have argued that this approach is so luminosity-hungry
that it is not practicable-there simply is not enough time or luminosity to study each 
threshold this way if, for example, low energy Supersymmetry is realized in nature.
We sought a resolution to this issue at Snowmass 2001 by posing a test case: can a 
realistic machine, in a finite amount of time, do justice to a physics-rich scenario?  
The paper describing this interesting exercise, and its answer in the affirmative, is 
included in the proceedings \cite{grannis-paper}, and is summarized here.

\subsection{Definition of the Scenario and Assumptions on Machine Performance}

\begin{figure}
\includegraphics[scale=0.5]{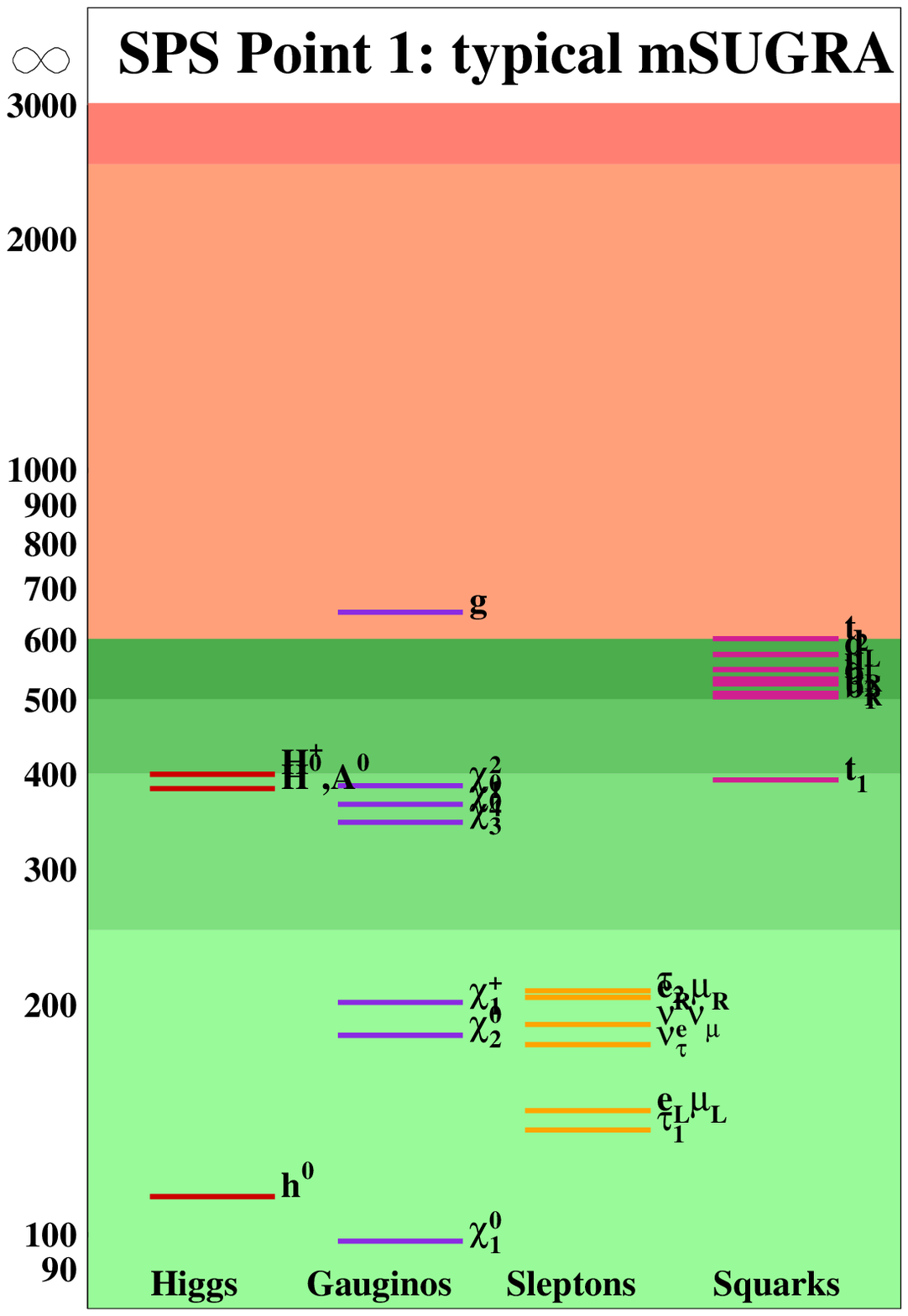}
\caption{The particle spectrum of the mSUGRA scenario considered for the 
definition of the run strategy.}
\label{fig:sps1}
\end{figure}

The new physics we imagined was a scenario rich with low mass supersymmetric 
particles and was one of the benchmarks defined for Snowmass studies.  The physics is 
defined by the following minimal Sugra parameters~\cite{Battaglia:2001zp}:  
$m_0$=100~GeV, $m_{1/2}$=250~GeV, $\tan \beta$=10, $A_0$=0, and $\mu$ positive.  
The resulting particle spectrum, and the 
principal decay modes of the particles, are given in Figure~\ref{fig:sps1}.  
The sleptons and 
sneutrinos are all kinematically accessible at 500~GeV in this scenario; the lighter 
chargino and the two lightest neutralinos can be pair produced at 500~GeV, while the 
heavier chargino and the two heaviest neutralinos can be made in association with the 
lighter states. The remaining particles in this scenario, with the exception of the 
light $h^0$ at 113~GeV, are inaccessible at LC-500: the lighter squarks and gluinos have 
masses between 500 and 600~GeV and the stop states have masses of 393 and 572~GeV. The 
remaining Higgs states are all near 380~GeV.

Machine performance was assumed to range from a few percent of design luminosity 
in the first year of LC operations up to the design value by year five,  allowing 
integration of 1000~fb$^{-1}$ over a seven year period. This integrated luminosity
would obtain if all the running were at 500~GeV, where the luminosity is highest.
For running at lower energies, the luminosity is scaled back in proportion to the 
energy, so the actual net integrated luminosity is lower than 1000~fb$^{-1}$ reflecting
some lower energy running.   It was assumed that electron beam polarization, of 
magnitude 80\%, was available, but that positron polarization was not. It was also 
assumed that $e^-e^-$ collisions could be arranged, albeit with luminosity reduced 
to 20\% of that possible for $e^+e^-$. Operation  with $\gamma\gamma$ collisions may 
also prove desirable, but it has not been considered here, as it would possibly be a 
later option.

The physics model defines the cross-sections for the pair and/or associated production 
of all sparticle states, which of course depend on the polarization of the incident 
electron beam.  
Cross-sections range from of order 10~fb for associated chargino and neutralino 
production and muon and tau sneutrino pair production, to roughly 50~fb
for smuon and stau pair production, to greater than 100~fb for
the lightest chargino and selecton pairs. 
Physics goals for the program included the full elucidation of the above sparticle 
spectrum, detailed studies of the Higgs, and a thorough study of  the top quark 
threshold region.

\subsection{Run Strategy}

Running at or near the highest energy was allotted about half the total integrated 
luminosity.  This ensured production of all the kinematically allowed states. The common
experimental signature for production of these SUSY states is lepton pairs plus
missing energy. Sparticle masses can be deduced by measuring the endpoints of the 
lepton energy distributions.  
In the case where a single sparticle is pair produced and decays via a two body process into a 
second sparticle and a lepton, a simple box energy 
spectrum emerges with unambiguous endpoints 
which are related to the masses of the parent and daughter sparticles.
The energy distributions are more complicated when several sparticles 
contribute to the same observable final state, leading to superpositions of 
such  spectra.  Papers cited in the Snowmass study have shown how these cases can be 
disentangled, and have provided the estimates of the precision with which sparticle 
masses can be obtained.  With appropriate scaling of the statistics, these results lead 
to estimates of how accurately end-point masses can be measured with the given 
luminosity assumptions.
Given this coarse determination of the masses of many sparticle states, one knows 
how to choose the machine energy for threshold scans of selected states, which can 
yield more accurate mass determinations. A simple, non-optimized scan strategy was 
employed; the luminosity is divided among 10 mini-runs which are distributed over energy
in even steps.  The scans yield some information on sparticle widths, in addition to 
precise information on the masses.
     Electron polarization is used to good advantage in the measurement process.  Using equal 
amounts of L and R running at the higher energy points helps distinguish L and R 
sparticles, even though they may decay to the same final states. For the threshold 
scans, the cross-sections for production of particular states can be enhanced by correct
choice of the electron polarization, so many scans use exclusively L or R electrons, 
but not both. Since the cross-section for $\tilde{e_R} \tilde{e_R}$ near threshold rises
as $\beta$ in $e^-e^-$ interactions (vs $\beta^3$ in $e^+e^-$), it is advantageous to 
use $e^-e^-$ collisions for the precision determination of the $\tilde{e_R}$ mass.
The resulting run plan is shown in Table~\ref{tab:runsps1}.  
Roughly one half of the running is 
at the nominal maximum energy of the machine, and a little is even above the nominal 
maximum (which is possible by trading luminosity for additional energy). The rest is 
distributed among threshold scans, with extra running time going to low cross section 
processes. A good deal of running is near the Higgs-strahlung threshold, so precision 
studies of the Higgs boson properties are possible.

\begin{table*}
\caption{\label{tab:runsps1}
Run allocations for the SPS1 Minimal Sugra parameters. 
}
%\begin{ruledtabular}
\begin{tabular}{|cccccl|}
\hline
Beams & Energy & Pol. & $\int {\cal L}dt$ 
  & [$\int {\cal L}dt]_{\rm equiv}$ & ~~~~~~~~Comments \\  \hline
$\ee $ & 500 & L/R & 335 & 335 & Sit at top energy for sparticle
       end point measurements \\ \hline
$\ee $ & 270 & L/R & 100 & 185 & Scan $\chz1~\chz2$ 
                                        threshold (R pol.) \\
       &     &     &     &     & Scan $\stau1~\stau1$ threshold 
                                        (L pol.) \\ \hline
$\ee $ & 285 & R & 50 & 85 & Scan $\smupr~\smumr$ threshold \\ \hline
$\ee $ & 350 & L/R & 40 & 60 & Scan $t\overline t$ threshold\\ 
       &     &     &     &     & Scan $\selr~\sell$ threshold 
                                      (L \& R pol.) \\
       &     &     &     &     & Scan $\chp1~\chm1$ threshold (L pol.) \\
                                        \hline
$\ee $ & 410 & L/R & 100 & 120 & Scan $\stau2~\stau2$ threshold \\ \hline
$\ee $ & 580 & L/R & 90 & 120 & Sit above $\chpm1~\chmp2$ threshold for 
                                   $\chpm2$ end point mass\\ \hline \hline
$e^-e^- $ & 285 & RR & 10 & 95 & Scan with $e^-e^-$ collisions for 
                   $\selr$ mass   \\ \hline
\end{tabular}
%\end{ruledtabular}
\end{table*}

\subsection{Results}

Table~\ref{tab:measmass} shows the accuracies of the mass determinations for both 
endpoint and threshold measurements. Most slepton, sneutrino, chargino, and light 
neutralino masses are determined with a fractional accuracy of a few parts per mil or 
better.  Only the staus and tau sneutrino are measured at the per cent level or worse 
(excepting $\chi^0_4$, which is observed but not well-measured). These are impressive 
precisions, which would add significantly to what will have been learned at the LHC.  
When these results are taken in conjunction with existing information from the LHC,
it may well be possible to distinguish the broad class of SUSY model responsible for the
observed spectral pattern.  In the context of the particular Sugra model, it will be 
possible to determine the fundamental model parameters to high precision.
\begin{table}
\caption{\label{tab:measmass}
Mass precision estimates in GeV for benchmark point SPS1 
for end point (EP), threshold scan (TH) and combined 
measurements, and the combined estimates for the RR1 point.}
%\begin{ruledtabular}
\begin{tabular}{|c|ccc|c|}
\hline
 ~ & ~ & SPS1 & ~ & RR1 \\
particle & $\delta M_{\rm EP}$ &
           $\delta M_{\rm TH}$ & $\delta M_{\rm SPS1}$ & 
             $\delta M_{\rm RR1}$ \\ \hline
$\selr$ & 0.19 & 0.02 & 0.02 & 0.02 \\
$\sell$ & 0.27 & 0.30  & 0.20 & 0.20 \\
$\smur$ & 0.08 & 0.13  & 0.07 & 0.13 \\
$\smul$ & 0.70 & 0.76  & 0.51 & 0.30 \\
$\stau1$ & $\sim 1 - 2$ & 0.64  & 0.64 & 0.85 \\
$\stau2$ & -- & 0.86  & 0.86 & 1.34 \\
$\snue$ & 0.23 & --  & 0.23 & 0.4 \\
$\snum$ & 7.0 & --  & 7.0 & 0.5 \\
$\snut$ & -- & --  & -- & 10.0 \\
$\chz1$ & 0.07 & --  & 0.07 & 0.07 \\
$\chz2$ & $\sim 1 - 2$ & 0.12  & 0.12 & 0.30 \\
$\chz3$ & 8.5 & --  & 8.5  & 0.30 \\
$\chz4$ & -- & --  & -- & observed \\
$\chpm1$ & 0.19 & 0.18  & 0.13 & 0.09 \\
$\chpm2$ & 4.1 & --  & 4.1 & 0.25 \\ \hline
\end{tabular}
%\end{ruledtabular}
\end{table}

The run scenario includes a lot of running above the $Z^0h^0$ threshold, so many 
Higgs parameters will be determined with high precision.  Many fundamental couplings 
will be measured to the level of 1 or 2\% for $g_{ZZh}$, $g_{WWh}$, $g_{bbh}$,  and 
$g_{\tau\tau h}$.  The $g_{cch}$ coupling will be measured to 4\% and the coupling to 
top will be measured to about 30\% from the top threshold running.  The Higgs mass is 
measured to 0.03\%; and its width is indirectly determined to 7\%.
The run plan devotes time to top threshold running.  This will fix the top mass to 
150~GeV and the width to 5\%.

The run time plan developed at Snowmass  provides a proof that a 1000~fb$^{-1}$ data 
set, gathered over a period of about seven years, is adequate for a full range of high 
precision measurements of sparticle masses, Higgs properties, and top properties in a 
physics-rich scenario.  A second mSugra point, considered in the {\sc Tesla} TDR 
(point RR1) 
and now excluded by the LEP-2 data, was also studied, to get some feel for the 
sensitivity of the conclusions to the particulars of the initial assumptions. The same 
conclusions hold.

\section{LC and Alternative Physics Scenarios}

Recently a whole new class of models has been considered.  
These models go under the general banner
``large extra dimensions'' or ``brane world.''  The basic idea is
that there are more than three spatial dimensions, and that the
geometry and size of the extra dimensions can help us explain the
huge differences in scales between, for example, the Planck scale
and the weak scale, and between the weak scale and the mass scale associated 
with neutrinos.
So far, complete models that explain the size and stability of the extra
dimension(s) are not available.  However, investigations into
these theoretical structures are relatively new  and complete models
may emerge given time.
It is already possible to identify generic signals that are expected
to be present in these models
Many contributed articles to these proceedings, and the chapters of
the Linear Collider Resource Book~\cite{nlc-resource} and the 
{\sc Tesla} TDR~\cite{tesla-tdr}, contain a great deal
of information about the phenomenology of extra dimensions.  In the following
we briefly list some broad categories of these theories and describe
how a linear collider could constrain them.

{\it New gravity dimensions:}
One of the most exciting recent ideas to address the question of
why $M_P/m_W\gg 1$ was introduced by Arkani-Hamed, Dimopoulos and
Dvali~\cite{Arkani-Hamed:1998rs}.  
They explain that if there are $n$ extra dimensions in a volume
$V_n=(2\pi R)^n$ then
the fundamental scale of gravity $M_D$ could be of order the weak scale
provided the radius of the extra dimensions satisfies the relation
$M_P^2=M_D^{2+n}R^n$.
The radius of compactification can be as large as $\sim1\, {\rm mm}$
for $n=2$ extra dimensions.

There are collider consequences in  such a model, the most
important of which involve gravitons,
which are more strongly coupled at higher energy than in ordinary gravity
because $M_D\simeq M_W$.  A classic signature
is $e^+e^-\to G^{(n)}\gamma$. The $G^{(n)}$ represent the Kaluza-Klein
excitations of the graviton kinematically accessible in the
collision and exit the detector without interacting. The result is a
final state of photon and missing energy.
Similar signatures arise from quarks annihilating
into gravitons plus jet at the LHC.  

\begin{figure}
\includegraphics[scale=0.45]{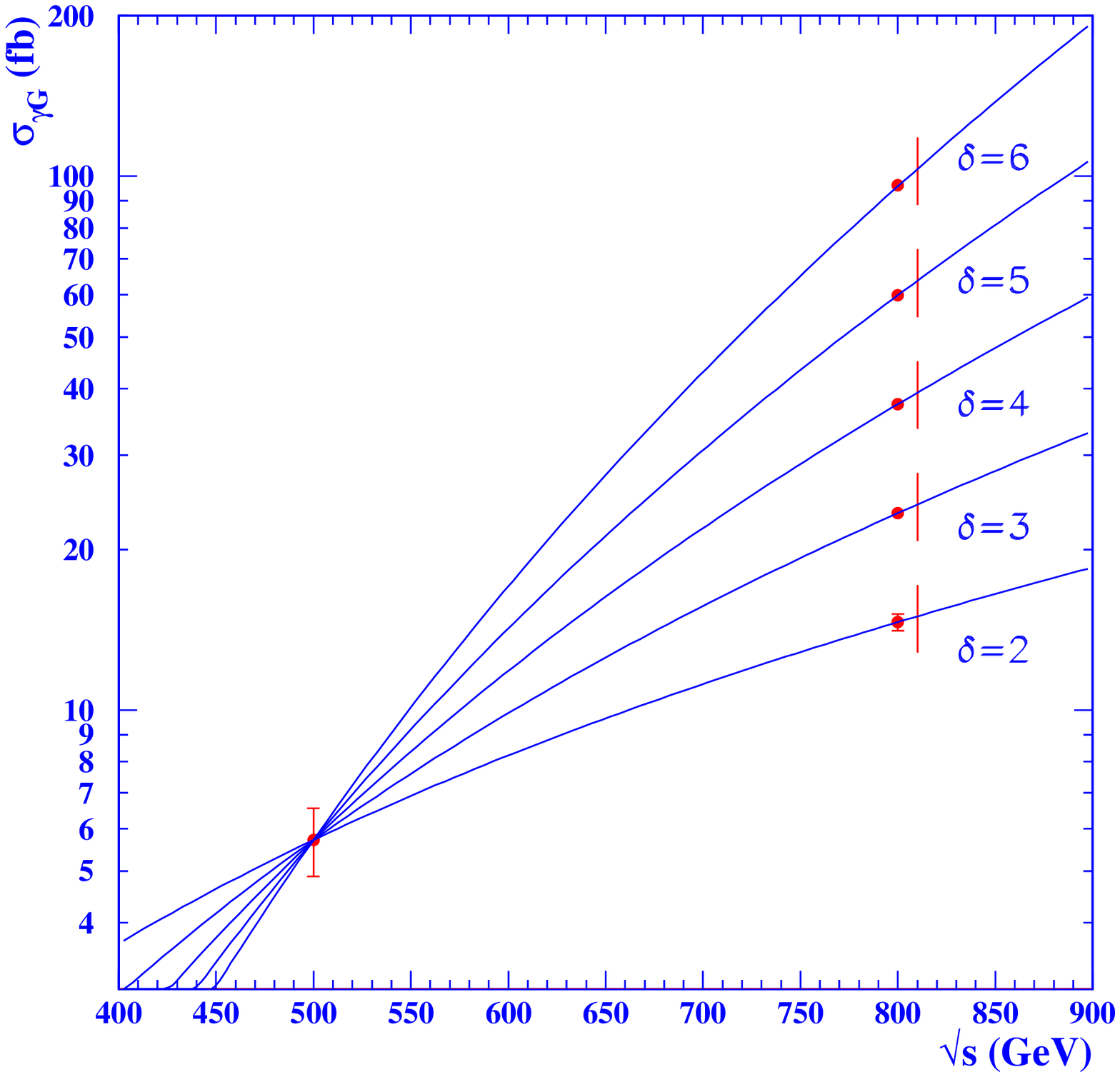}
\caption{Determination of the number of extra dimensions from anomalies in the 
single photon cross section measured at $\sqrt{s}$=500~GeV and 800~GeV. Points 
with error bars indicate the expected measurement performed on 500~fb$^{-1}$ of 
data with polarized electron and positron beams. The curves show the expected 
dependence of the cross section on the collision energy for different number of 
extra dimensions (from~\cite{tesla-tdr}).}
\label{fig:e3001_fig13}
\end{figure}

The production cross-section depends strongly on the center of mass
energy, and scales with energy as a power of the number of extra
dimensions, $\sigma \sim s^{n/2}/M_D^{2+n}$.  Therefore it is possible
to vary the center of mass energies and fit for the number of
extra dimensions by seeing the energy scaling behavior of the cross-section.
This is demonstrated nicely in Figure~\ref{fig:e3001_fig13}.

In this class of models the gauge fields of the standard model cannot
propagate in the extra dimensions. If the size of the extra
dimension is ${\rm TeV}^{-1}$, then 
gauge fields may propagate in the  extra 
dimensions~\cite{Antoniadis:1990ew}.  This  possibility
uses the compacification scale to to set a scale for electroweak symmetry breaking 
(or supersymmetry breaking, {\it etc.}). 
For example, consider all gauge fields living in
$4+1$ spatial dimensions, where the fifth dimension is compactified
in a volume characterized by a radius $R$.   The gauge fields will then
all have Kaluza-Klein excitations at integer multiples of $1/R$.  These
KK states have couplings determined by  their zero-mode counterparts
and are therefore predicted.

The search for KK excitations of gauge bosons is similar to the 
searches for $Z'/W'$ bosons in theories with extra gauge symmetries.
The strategies of those searches can be directly applied to
KK searches.  While it may not be possible to produce the states
directly, there presence will cause distortions in, for example,
$e^+e^-\to \mu^+\mu^-$ and a  reach well beyond the center of mass
energy can be obtained.
   For  example, at a 500 GeV linear collider
with $100\, {\rm fb}^{-1}$ integrated luminosity, KK states cause clear
deviations from the SM predictions of $e^+e^- \to f\bar f$ if the
KK masses are below 10 TeV. 

%The close connection to $Z'/W'$ search strategies with KK search strategies,
%also means that there is a risk of confusing the two once a signal
%is detected.  It has been shown 
%that a careful combination of both 
%LHC data and LC data is useful to distinguish between the 
%two~\cite{Rizzo:1999en,Rizzo:2001jj} providing
%another example of the indispensible role a linear collider would play
%in understanding the big questions we will be facing in the presence
%of a signal.

Another class of models invokes {\it Warped gravity dimensions.}
Here  it has been shown
that the weak  scale may arise as a byproduct 
of a warped-geometry extra-dimensional space~\cite{Randall:1999ee}.  
This possibility and many
variants of it have phenomenological predictions, the most spectacular
of which is the observation of TeV-scale KK excitations
of the graviton (or other states), whose masses are spaced  in a
pattern characteristic of this class of models.
One of the earliest signs of warped extra dimensions would be
the discovery of the radion -- a modulus associated with the stability
of the extra dimensions. Models of brane separation stabilization 
imply that the radion is expected to be the lightest state directly
related to the extra dimension.  Recent papers have emphasized how similar
the radion is to Higgs in collider searchers, but the branching fractions
follow a decidedly different pattern~\cite{Kribs:2001ic}. 
Careful studies of the 
production cross-sections and branching
fractions of the radion and Higgs boson may even provide
information about gravity-induced mixing between the radion and Higgs boson.

%At colliders of much higher energy, other manifestations of extra
%dimensions may be visible. If the underlying theory is a string theory
%then it may be possible to produce black holes. 
%The cross-section can be very large and the
%signature spectacular. A multi-TeV collider will also be able to make
%detailed studies of the high mass $WW$ system from the process
%$e^+e^-\to WW\nu\nu$ and probe models of electro-weak symmetry
%breaking that involve as strongly coupled gauge boson sector. Such a
%collider would also have access to the heavier states in a
%supersymmetric model such as the heavy higgs bosons and possibly squarks. 

Multi-TeV $e^+e^-$ collisions are expected to break new ground and extend
the search for new physics well beyond the LHC reach~\cite{Barklow:2001mm}. 
In particular precision measurements will allow to probe scales of
new physics in the range of several hundred TeV.
%, the ultimate region
%in which the SM with a light Higgs could escape  from new physics, but 
%where it finally will have to  reveal its  signs.

\begin{figure}
\begin{tabular}{c c}
\includegraphics[scale=0.45]{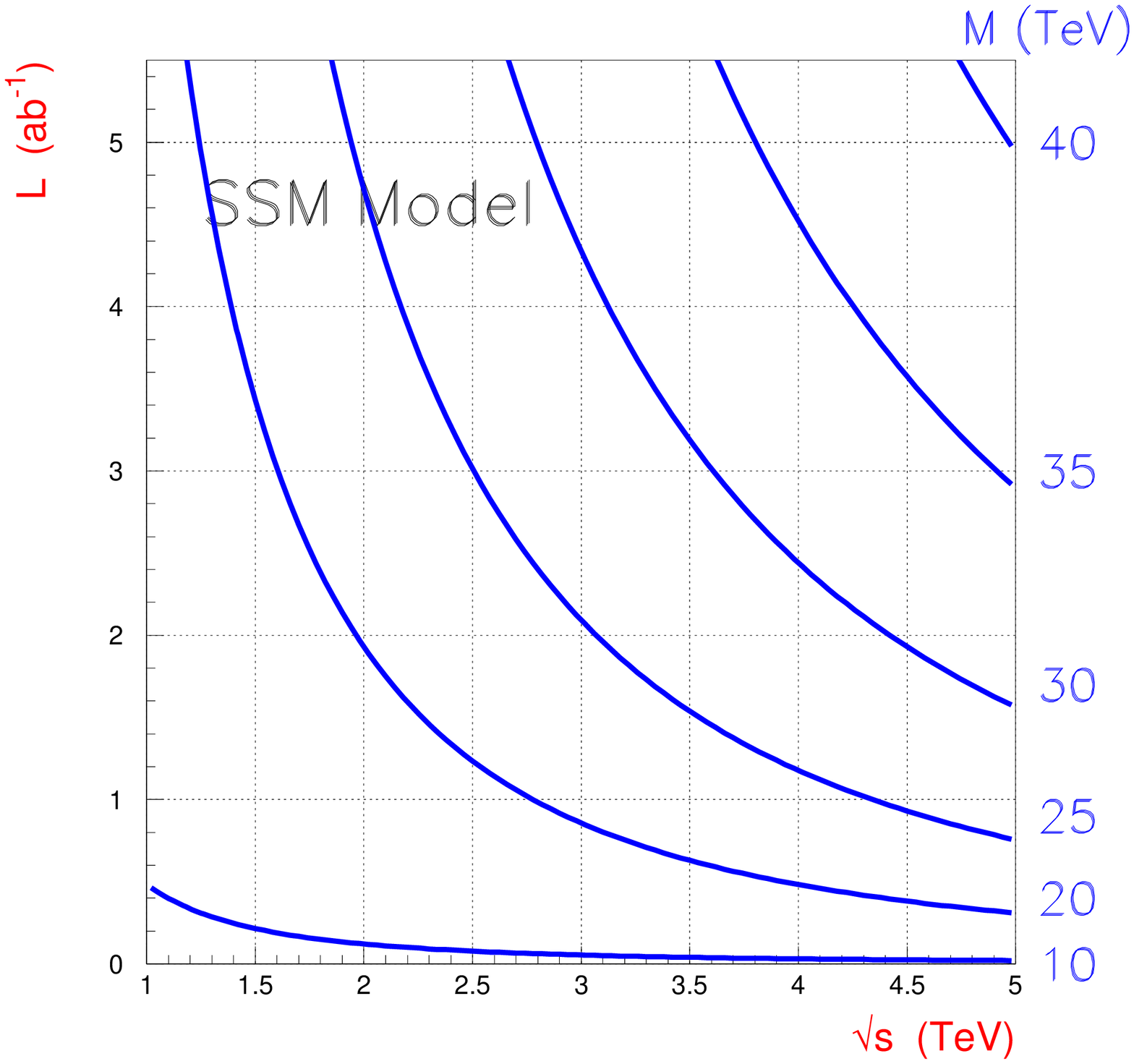} &
\includegraphics[scale=0.41]{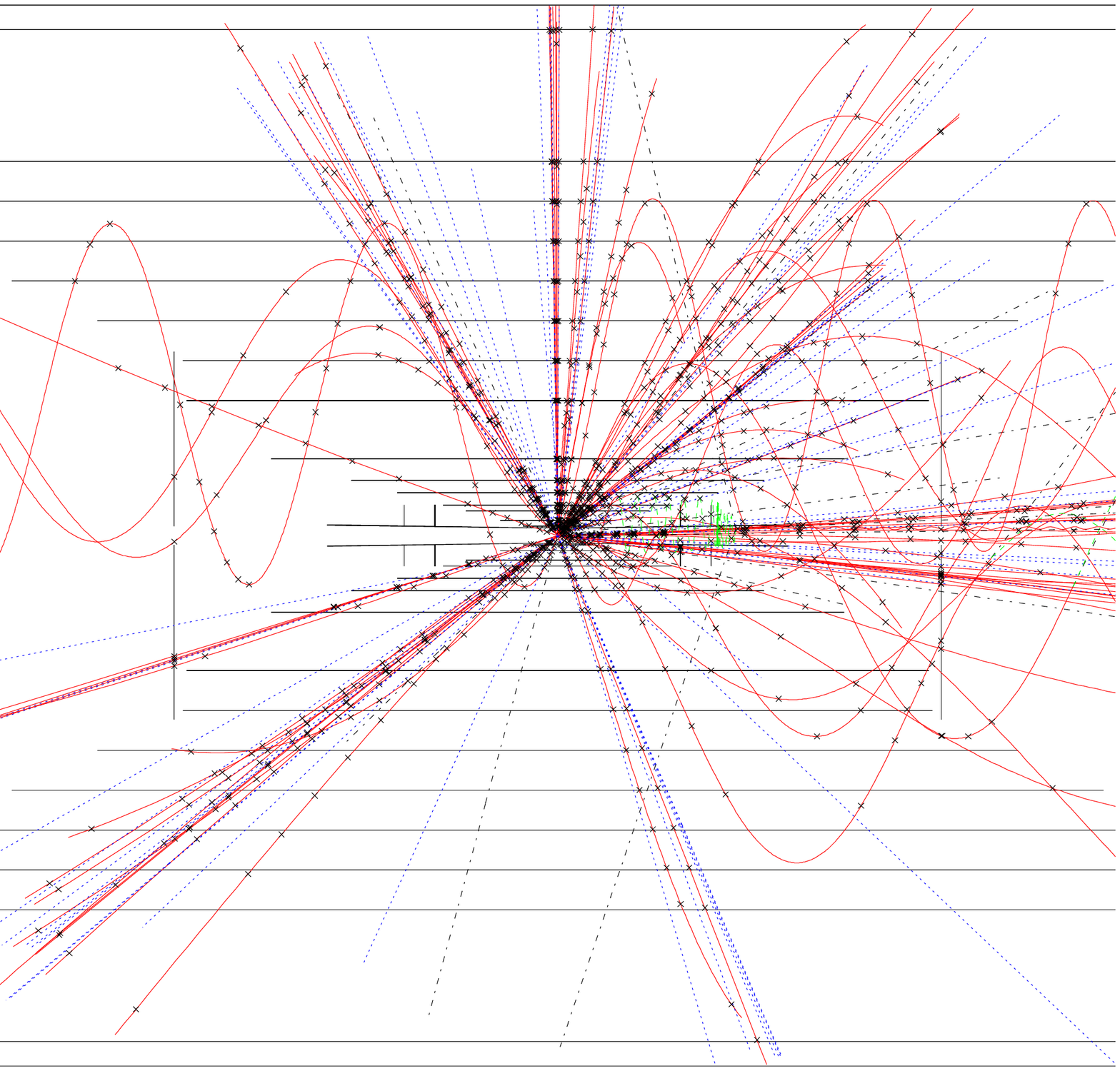} \\
\end{tabular}
\caption{Examples of multi-TeV physics. Left: the sensitivity to an additional 
$Z'$ gauge boson through measurements of electro-weak observables as a 
function the integrated luminosity and center-of-mass energy 
(from~\cite{Battaglia:2001xx}). 
Right: an event of production and evaporation of a black-hole in 
5~TeV $e^+e^-$ collisions (from~\cite{Barklow:2001mm}).}
\label{fig:e3001_fig14}
\end{figure}

Several possible scenarios of new physics in the TeV region are being considered.  
The upper end of the SUSY sparticle mass spectrum is likely to extend up to, 
or beyond, 1~TeV, requiring multi-TeV $e^+e^-$ collisions to produce those 
particles and  measure their properties with high precision. 
Beyond 1~TeV, the phenomenology implied by alternative physics scenarios,
such as new gauge bosons, Kaluza-Klein excitations, a new strong
sector, { \it etc.}, 
leads in many cases to  the production of new resonances in the multi-TeV range. 
The total production cross section as a function of  center of mass energy could 
therefore show a spectacular new structure.
For a  collider with large enough energy, {\it e.g.} the production and decay of 
KK excitations of the graviton can give direct access to the graviton 
self couplings~\cite{Battaglia:2001id}. 
If the center-of-mass energy is even larger than  the fundamental scale of gravity, 
which may happen in scenarios with extra dimensions, $e^+e^-$
collisions may even result in the abundant production of black 
holes~\cite{Giddings:2001ih,bh}, which would subsequently
evaporate in a large number of partons (see Figure~\ref{fig:e3001_fig14}), and 
re-shape our view of the evolution of elementary processes at high energies.

\section{LC Extendibility, Upgradeability and multi-TeV Option}

While the physics program of a linear collider providing collisions at energies in 
the range 0.3~TeV $<\sqrt{s}<$ 0.5~TeV appears rich and compelling, its extension to 
higher energies must already be considered~\cite{dq4-report}. 
The physics case for this extension, 
cannot be fully detailed from our present understanding and it will
most likely be specified one the signals and measurements obtained
from the data from  the {\sc LHC}  and 
also from  the LC operation at 0.5~TeV and  lower. 
 However, our present favored picture of the new physics possibilities and even 
the alternative physics scenarios discussed in the previous section, all will 
eventually require study with $e^+e^-$ collisions at energy beyond 0.5~TeV.  
The investigation of the coupling of the Higgs with the heaviest known elementary 
particle can only be performed at energies of 0.8~TeV-1.0~TeV where the radiation 
of a Higgs boson off a top quarks measures the top Yukawa coupling $g_{Htt}$.
If SUSY exists, the spectrum of supersymmetric partners
may already be studied, in part, at lower energies but it will most likely require 
energies in excess of 0.5~TeV to make these studies exhaustive. 
The approach to  1~TeV also promises to open an additional window on 
new physics. If the breaking of the electro-weak symmetry is not due to the Higgs 
mechanism, then the $e^+e^- \rightarrow W^+W^-\nu\bar{\nu}$ and $Z^0Z^0\nu\bar{\nu}$ 
processes will reveal a new dynamics of gauge boson interactions~\cite{Barklow:2001is}. 
Even if the Higgs 
boson measurements has  validated the Higgs mechanism, the study of triple gauge 
boson couplings via pair production 
at energies beyond 0.5~TeV will be needed to look for the effect of 
new interactions~\cite{tesla-tdr,Battaglia:2001nn}.

\begin{figure}[t]
\includegraphics[scale=0.45]{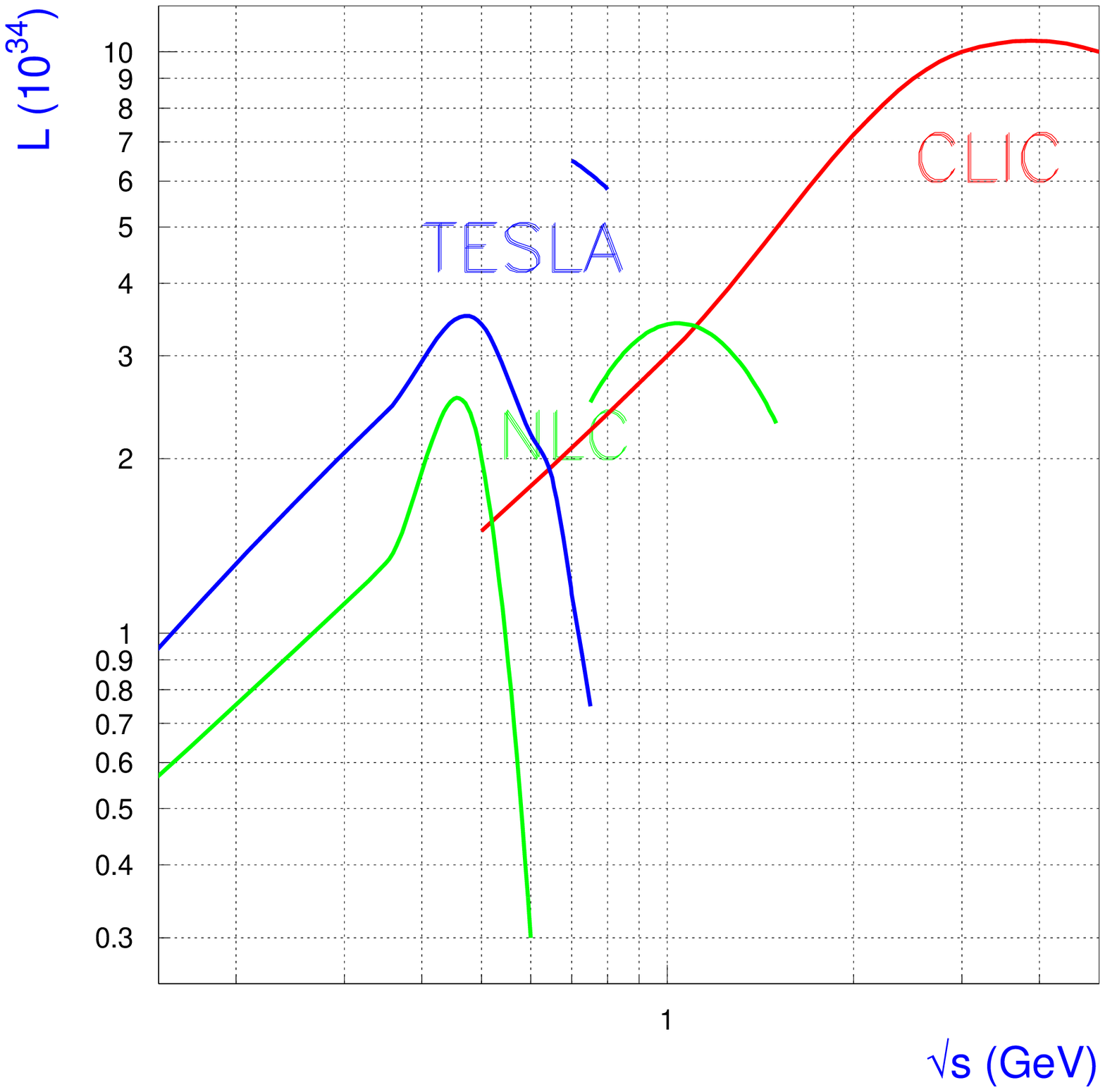} 
\caption{Estimated luminosity as a function of the $\sqrt{s}$ energy for
different LC projects. The two sets of curves for NLC and {\sc Tesla} 
refer to the baseline projects where the luminosity is traded for the beam energy, 
keeping the beam power constant, and to a possible linac upgrade.}
\label{fig:e3001_fig15}
\end{figure}

The extension of the LC energy can be achieved by three different
strategies (see Figure~\ref{fig:e3001_fig15}). 
The first is to increase the energy at the expense of the luminosity, where 
the limit is set by the available power.
The second is to increase the accelerating gradient and/or adiabatically extend the 
active linac by adding extra accelerating structure. This scheme was successfully adopted
at {\sc LEP} in raising the collision energy by more than a factor of two. Here the 
limit is set by the site length, the achieved gradient and the RF power. Beyond these 
limits, a further energy upgrade could be achieved by replacing the full active part of 
the linac. 

In the case of an X-band LC, the first scenario is expected to provide $e^+e^-$ 
collisions at $\sqrt{s}$ = 0.60~TeV with about 15\% of the nominal luminosity at 
0.5~TeV. The {\sc Tesla} superconducting technology can achieve collisions at 
$\sqrt{s}$ = 0.65 (0.75)~TeV with about 55 (10)\% of the nominal luminosity at 0.5~TeV. 
This may be of interest for the first exploration of this higher energy region or for 
clarifying the nature of possible new signals which have emerged at the lower energies. 
However in
order to fully profit from the higher energy, the luminosity also needs to be increased, 
to compensate for the fall of s-channel cross sections. This is addressed in the 
different
designs, that foresee a possible second stage upgrade. The {\sc NLC} project aims at 
achieving 1~TeV with luminosity of 3.4$\times 10^{34}$cm$^{-2}$s$^{-1}$, by doubling the
number of components and, possibly, increasing the gradient.
The {\sc Tesla} TDR proposes an upgrade, based on operating the linac with a gradient of
35~MeV/m, instead of 23.4~MeV/m. This would enable collisions 
at $\sqrt{s}$ = 0.8~TeV with a luminosity of $5.8 \times 10^{34}$ cm$^{-2}$s$^{-1}$.
Both these upgrade paths require accelerating gradients higher than presently 
demonstrated and significant R\&D is presently being pursued. 
However, the first indications are quite encouraging. 

Beyond these energies, the extensions of the SC and X-band technology are more 
speculative. In order to attain collisions at energies in excess of 1~TeV, with 
high luminosity, significantly higher gradients are necessary so that  the total 
linac length and the needed power  are within acceptable limits. Also, the number of 
active elements in the linac must be kept lowe enough to ensure reliable operations.
The two-beam acceleration scheme, developed at CERN in the framework of the 
{\sc Clic} study in collaboration with other laboratories in Europe and in the US, aims 
at a collider capable of providing collisions at $\sqrt{s}$ = 1~TeV - 5~TeV with a 
luminosity of $1 \times 10^{35}$~cm$^{-2}$s$^{-1}$ at 3~TeV~\cite{clic}. 
To demonstrate the 
feasibility of the {\sc Clic} scheme and the achievability of the design gradient of 
150~MeV/m, 
an intensive program of tests is under way.
Whether a multi-TeV LC is best realized as a standalone facility or as the second stage 
of an existing TeV-class collider must be considered before committing
to a particular site and a layout. 
In fact, in the latter case the possibility to install the two linacs at 
a finite crossing angle, which is necessary to avoid parasitic collisions upstream of 
the interaction region, and the site requirements in terms of ground stability and 
coherence, which is needed to preserve luminosity from the collision of sub-nanometer 
size beams, both need to be taken into account.

Beyond about 1~TeV the fusion processes start becoming the dominant modes of fermion and
boson pair production. From 1~TeV to 3~TeV, the production of $WW\nu\nu$, $ZZ\nu\nu$ 
and $t\bar{t}\nu\nu$ starts exceeding that from the $s$-channel $WW$, $ZZ$ and $ZZ$ 
processes. Since these cross sections
increase as $\log(s/M^2)$, the luminosity can be traded with energy in this regime.
In some searches, the option of colliding photons at a Compton collider extends the reach
in mass. An interesting example is offered by heavy Higgs bosons, that are pair produced
in $e^+e^-$ collisions only up to $M_H \simeq 0.5 \times \sqrt{s}$ and may be observed 
in $\gamma\gamma$ collisions up to $M_H \simeq 0.8 \times \sqrt{s_{ee}}$. 

At multi-TeV energies, the phenomenology implied by alternative physics scenarios, such 
as the existence of extra dimensions, may yield spectacular signatures. Large resonances 
corresponding to the KK excitations of gauge bosons or new gauge bosons may design a 
different landscape for the evolution of the cross sections of two fermion processes. 
At the same time it is important to ensure a good sensitivity to the scaling of 
electro-weak observables at these high energy. 
Measurements of two fermion cross sections and asymmetries with good statistical 
accuracy may represent a unique window on new phenomena at mass scales 
of the order of tens or hundreds of TeV~\cite{Battaglia:2001xx}. 
As the typical sensitivity to these new physics 
scales with energy and integrated luminosity as $(s \times {\cal{L}})^{1/4}$, trade-offs 
between these two fundamental parameters can be defined 
(see Figure~\ref{fig:e3001_fig14}).

\section{Conclusions}

The constraints from  precision tests of the standard model provide
convincing evidence that the Higgs boson of the standard model should
be light \cite{lepewwg:2001}. These tests can be reinterpreted
in the context of extensions to the standard model and imply~\cite{Peskin:2001rw} 
that the physics responsible for electroweak symmetry breaking, or 
a manifestation of new physics, will be accessible to a linear collider 
with center of mass energy of 500~GeV. The LHC will
investigate this energy regime before a linear collider could begin
operation.
While the LHC will enable many measurements of the properties of a
Higgs boson, data from a linear collider will be needed to measure the
properties in detail and complete the study; in particular some decay modes are expected
to be inaccessible at LHC. The need to measure the Higgs properties with high precision 
at a linear collider demand a detector that is capable of fully exploiting
the physics. This demands, in particular, a vertex system capable of discriminating
jets with charm and bottom quarks, excellent track momentum resolution and good 
jet energy determination.

The standard model is known to be incomplete. Many extensions have
been proposed. In the case of supersymmetry, a linear collider is
capable of precision measurements of the masses and other properties of the
supersymmetric particles that are kinematically accessible. Detailed 
measurements of
standard model processes such as $e^+e^-\to WW$ can be used to
constrain new physics models. Such measurements can be used, in conjunction with those 
made at the LHC, to give a more complete understanding of the underlying physics.
A linear collider of high luminosity
also offers the opportunity to extend the current precision tests of
the standard model by obtaining a sample of $10^9$ $Z$ bosons.

An enormous amount of accelerator R\&D has been undertaken over the
last decade (for more informations see~\cite{m3report}). 
It is clear that this work
is sufficiently mature that the construction of a collider of energy
500~GeV could be undertaken. R\&D should be pursued to guarantee high gradient 
operation in the accelerating structures, so that an eventual upgrade to higher 
energies is possible. Electron polarization is a vital component in the initial 
operation of a linear collider. R\&D should be pursued with a view to providing 
psoitron polarization. While such polarization need not be
part of the initial configuration, it is highly desirable for a complete
exploitation of the physics, if supersymmetric particles are found or
if the collider is used to obtain a large sample of $Z$ bosons.
The ability to extend the initial energy to of order 1~TeV should be 
an essential part of the facility.

\begin{acknowledgments}
The work was supported in part by the Director, Office of Energy Research, Office of 
High Energy and Nuclear Physics of the U.S. Department of Energy under Contracts 
DE-AC03-76SF00098 and DE-AC02-98CH108886. Accordingly, the U.S. Government retains a 
nonexclusive, royalty-free license to publish or reproduce the published form of this
contribution, or allow others to do so, for U.S. Government purposes.
\end{acknowledgments}

\section{Appendix -- Charge to E3 group}

\begin{itemize}

\item The eeLC group should coordinate with the physics groups to help compile and critically examine the case
        for an initial phase of the ee collider at a cm energy of up to about 500 GeV, depending on the results from
        prior experiments at the Tevatron and LHC. For some representative physics scenarios, what is a reasonable
        goal for integrated luminosity at various cm energies, beam polarizations and beam particles? is there a
        compelling initial physics program at a luminosity of a few
        $\times 10^{33}$cm$^{-2}$s${-1}$? Are there particular advantages
        or challenges to experimentation raised by the different running conditions in the {\sc Tesla} and NLC/JLC
        designs?  
     \item The eeLC group should review the case for and feasibility
       of special options for LC operations:
\begin{itemize}  
           \item Catalog the physics needs that may require positron polarization, gamma gamma collisions or e-e-
             collisions. Compare the capabilities of an electron-positron collider and a gamma-gamma collider for
             making detailed measurements of the properties of Higgs bosons, and for discovering Higgs bosons.
             What are the R\&D issues remaining for each option? What are the requirements on the initial design to
             allow any of these to be added after the initial phase?  
          \item Examine the case for high-luminosity operation at the Z pole. What are the benefits and drawbacks
             from the design of a special beam delivery system for low energy collisions? Should there be a
             special detector devoted to operating below 500 GeV?  
           \item  What special requirements are imposed if a free electron laser program is added to the high energy
             physics facility? What should the HEP community do to
             facilitate the potential for a FEL program?  
\end{itemize}
     \item Evaluate the scientific case for an initial-phase "Higgs factory" at an energy of about 300 GeV.  
      \item What new physics landmarks come into view as the energy of a linear collider is raised to 1 TeV; to 1.5 TeV;
        to 2 TeV; to 5 TeV? What luminosity and other performance characteristics would be required to maximize the
        scientific output?  
     \item Are there particular issues that detector R\&D must address to guarantee the productivity of a linear collider? 
     \item What are the beam physics limits and accelerator limits imposed on LC performance, and what are the
        primary outstanding R\&D issues that are critical to study in the next several years?  
     \item The eeLC group should assume that a technical review panel will likely be established within the next year
        to evaluate the superconducting L-band and warm rf X- or C-band accelerator proposals. That review,
        conducted under the auspices of some worldwide body, would examine the performance parameters of the
        machines, the technical risks, needed R\&D, comparative costs and upgradability. Without undertaking the
        work that such a panel would do, the eeLC group should work to sharpen the questions that this review
        panel should examine, and consider the way in which the panel should operate.  
     \item What are the paths for upgrade of an initial LC, both in energy and in luminosity? What extensions in energy
        using the original {\sc Tesla} or X-band LC designs are feasible? What R\&D issues should be given priority?
        What is the possibility of upgrading either {\sc Tesla} or 
X-band LC using two-beam drive power sources? What
        are the critical R\&D issues? What constraints on the initial phase would ultimate conversion to two-beam
        drive impose?  
     \item How can full international collaboration on a LC project be realized? Is it feasible to assign full responsibility
        for design, construction, commissioning, test and operation of major subsystems to different portions of the
        world community while maintaining effective overall project
        management?
\end{itemize}

\bibliography{e3_summary}

\begin{thebibliography}{57}
\expandafter\ifx\csname natexlab\endcsname\relax\def\natexlab#1{#1}\fi
\expandafter\ifx\csname bibnamefont\endcsname\relax
  \def\bibnamefont#1{#1}\fi
\expandafter\ifx\csname bibfnamefont\endcsname\relax
  \def\bibfnamefont#1{#1}\fi
\expandafter\ifx\csname citenamefont\endcsname\relax
  \def\citenamefont#1{#1}\fi
\expandafter\ifx\csname url\endcsname\relax
  \def\url#1{\texttt{#1}}\fi
\expandafter\ifx\csname urlprefix\endcsname\relax\def\urlprefix{URL }\fi
\providecommand{\bibinfo}[2]{#2}
\providecommand{\eprint}[2][]{\url{#2}}

\bibitem[{\citenamefont{{The LEP collaborations and the SLD heavy flavour and
  electro-weak group}}(2001)}]{lepewwg:2001}
\bibinfo{author}{\bibnamefont{{The LEP collaborations and the SLD heavy flavour
  and electro-weak group}}}  (\bibinfo{year}{2001}),
  \bibinfo{note}{\uppercase{LEPEWWG/2001-01}}.

\bibitem[{\citenamefont{{The LEP Collaborations and the LEP Working Group for
  Higgs boson searches}}(2001)}]{lephwg:2001}
\bibinfo{author}{\bibnamefont{{The LEP Collaborations and the LEP Working Group
  for Higgs boson searches}}}  (\bibinfo{year}{2001}),
  \bibinfo{note}{\uppercase{CERN-EP/2001-055}}, \eprint{hep-ex/0107029}.

\bibitem[{\citenamefont{Schumm}(2001)}]{schumm}
\bibinfo{author}{\bibfnamefont{B.}~\bibnamefont{Schumm}}
  (\bibinfo{year}{2001}), \bibinfo{note}{in these proceedings}.

\bibitem[{\citenamefont{Battaglia et~al.}(2001{\natexlab{a}})}]{grannis-paper}
\bibinfo{author}{\bibfnamefont{M.}~\bibnamefont{Battaglia}}
  \bibnamefont{et~al.} (\bibinfo{year}{2001}{\natexlab{a}}), \bibinfo{note}{in
  these proceedings}.

\bibitem[{\citenamefont{Barklow and De~Roeck}(2001)}]{Barklow:2001mm}
\bibinfo{author}{\bibfnamefont{T.~L.} \bibnamefont{Barklow}} \bibnamefont{and}
  \bibinfo{author}{\bibfnamefont{A.}~\bibnamefont{De~Roeck}}
  (\bibinfo{year}{2001}), \bibinfo{note}{in these proceedings},
  \eprint[http://arXiv.org/abs]{hep-ph/0112313}.

\bibitem[{\citenamefont{Burrows and Patterson}(2001)}]{dq4-report}
\bibinfo{author}{\bibfnamefont{P.}~\bibnamefont{Burrows}} \bibnamefont{and}
  \bibinfo{author}{\bibfnamefont{R.}~\bibnamefont{Patterson}}
  (\bibinfo{year}{2001}), \bibinfo{note}{in these proceedings}.

\bibitem[{\citenamefont{Erler et~al.}(2001)}]{Erler:2001su}
\bibinfo{author}{\bibfnamefont{J.}~\bibnamefont{Erler}} \bibnamefont{et~al.}
  (\bibinfo{year}{2001}), \bibinfo{note}{in these proceedings},
  \eprint[http://arXiv.org/abs]{hep-ph/0112070}.

\bibitem[{\citenamefont{Velasco et~al.}(2001)}]{Velasco:2001vg}
\bibinfo{author}{\bibfnamefont{M.~M.} \bibnamefont{Velasco}}
  \bibnamefont{et~al.}  (\bibinfo{year}{2001}), \bibinfo{note}{in these
  proceedings}, \eprint[http://arXiv.org/abs]{hep-ex/0111055}.

\bibitem[{\citenamefont{Aguilar-Saavedra et~al.}(2001)}]{tesla-tdr}
\bibinfo{author}{\bibfnamefont{J.~A.} \bibnamefont{Aguilar-Saavedra}}
  \bibnamefont{et~al.}, \emph{\bibinfo{title}{\uppercase{TESLA}
  \uppercase{T}echnical \uppercase{D}esign \uppercase{R}eport}}
  (\bibinfo{year}{2001}), \bibinfo{note}{\uppercase{DESY-2001-011}},
  \eprint[http://arXiv.org/abs]{hep-ph/0106315}.

\bibitem[{\citenamefont{Abe et~al.}(2001{\natexlab{a}})}]{nlc-resource}
\bibinfo{author}{\bibfnamefont{T.}~\bibnamefont{Abe}} \bibnamefont{et~al.},
  \emph{\bibinfo{title}{Linear collider physics resource book for
  \uppercase{S}nowmass 2001}} (\bibinfo{year}{2001}{\natexlab{a}}),
  \bibinfo{note}{\uppercase{SLAC-570}},
  \eprint[http://arXiv.org/abs]{hep-ph/0106055-058}.

\bibitem[{\citenamefont{Abe et~al.}(2001{\natexlab{b}})}]{Abe:2001gc}
\bibinfo{author}{\bibfnamefont{K.}~\bibnamefont{Abe}} \bibnamefont{et~al.},
  \emph{\bibinfo{title}{Particle physics experiments at \uppercase{JLC}}}
  (\bibinfo{year}{2001}{\natexlab{b}}),
  \bibinfo{note}{\uppercase{KEK-REPORT-2001-11}},
  \eprint[http://arXiv.org/abs]{hep-ph/0109166}.

\bibitem[{\citenamefont{{The LEP Collaborations and the LEP W Working
  Group}}(2001)}]{lepw:2001}
\bibinfo{author}{\bibnamefont{{The LEP Collaborations and the LEP W Working
  Group}}}  (\bibinfo{year}{2001}),
  \bibinfo{note}{\uppercase{LEPEWWG/MASS/2001-02}}.

\bibitem[{\citenamefont{Affolder et~al.}(2001)}]{cdfw}
\bibinfo{author}{\bibfnamefont{T.}~\bibnamefont{Affolder}} \bibnamefont{et~al.}
  (\bibinfo{collaboration}{CDF}), \bibinfo{journal}{Phys. Rev.}
  \textbf{\bibinfo{volume}{D64}}, \bibinfo{pages}{052001}
  (\bibinfo{year}{2001}), \eprint[http://arXiv.org/abs]{hep-ex/0007044}.

\bibitem[{\citenamefont{Abazov et~al.}(2001)}]{d0w}
\bibinfo{author}{\bibfnamefont{V.~M.} \bibnamefont{Abazov}}
  \bibnamefont{et~al.} (\bibinfo{collaboration}{D0})  (\bibinfo{year}{2001}),
  \eprint[http://arXiv.org/abs]{hep-ex/0106018}.

\bibitem[{\citenamefont{Abe et~al.}(1999)}]{cdftop}
\bibinfo{author}{\bibfnamefont{F.}~\bibnamefont{Abe}} \bibnamefont{et~al.}
  (\bibinfo{collaboration}{CDF}), \bibinfo{journal}{Phys. Rev. Lett.}
  \textbf{\bibinfo{volume}{82}}, \bibinfo{pages}{271} (\bibinfo{year}{1999}),
  \eprint[http://arXiv.org/abs]{hep-ex/9810029}.

\bibitem[{\citenamefont{Abachi et~al.}(1995)}]{d0top}
\bibinfo{author}{\bibfnamefont{S.}~\bibnamefont{Abachi}} \bibnamefont{et~al.}
  (\bibinfo{collaboration}{D0}), \bibinfo{journal}{Phys. Rev. Lett.}
  \textbf{\bibinfo{volume}{74}}, \bibinfo{pages}{2632} (\bibinfo{year}{1995}),
  \eprint[http://arXiv.org/abs]{hep-ex/9503003}.

\bibitem[{\citenamefont{Peskin and Wells}(2001)}]{Peskin:2001rw}
\bibinfo{author}{\bibfnamefont{M.~E.} \bibnamefont{Peskin}} \bibnamefont{and}
  \bibinfo{author}{\bibfnamefont{J.~D.} \bibnamefont{Wells}},
  \bibinfo{journal}{Phys. Rev.} \textbf{\bibinfo{volume}{D64}},
  \bibinfo{pages}{093003} (\bibinfo{year}{2001}),
  \eprint[http://arXiv.org/abs]{hep-ph/0101342}.

\bibitem[{\citenamefont{Kane et~al.}(1993)\citenamefont{Kane, Kolda, and
  Wells}}]{Kane:1993kq}
\bibinfo{author}{\bibfnamefont{G.~L.} \bibnamefont{Kane}},
  \bibinfo{author}{\bibfnamefont{C.}~\bibnamefont{Kolda}}, \bibnamefont{and}
  \bibinfo{author}{\bibfnamefont{J.~D.} \bibnamefont{Wells}},
  \bibinfo{journal}{Phys. Rev. Lett.} \textbf{\bibinfo{volume}{70}},
  \bibinfo{pages}{2686} (\bibinfo{year}{1993}),
  \eprint[http://arXiv.org/abs]{hep-ph/9210242}.

\bibitem[{\citenamefont{Espinosa and Quiros}(1998)}]{Espinosa:1998re}
\bibinfo{author}{\bibfnamefont{J.~R.} \bibnamefont{Espinosa}} \bibnamefont{and}
  \bibinfo{author}{\bibfnamefont{M.}~\bibnamefont{Quiros}},
  \bibinfo{journal}{Phys. Rev. Lett.} \textbf{\bibinfo{volume}{81}},
  \bibinfo{pages}{516} (\bibinfo{year}{1998}),
  \eprint[http://arXiv.org/abs]{hep-ph/9804235}.

\bibitem[{\citenamefont{Martinez}(2001)}]{martinez}
\bibinfo{author}{\bibfnamefont{M.}~\bibnamefont{Martinez}}
  (\bibinfo{year}{2001}), \bibinfo{note}{in these proceedings}.

\bibitem[{\citenamefont{Hoang et~al.}(2000)}]{Hoang:2000yr}
\bibinfo{author}{\bibfnamefont{A.~H.} \bibnamefont{Hoang}}
  \bibnamefont{et~al.}, \bibinfo{journal}{Eur. Phys. J. direct}
  \textbf{\bibinfo{volume}{C3}}, \bibinfo{pages}{1} (\bibinfo{year}{2000}),
  \eprint[http://arXiv.org/abs]{hep-ph/0001286}.

\bibitem[{\citenamefont{Xella~Hansen}(2001)}]{hansen}
\bibinfo{author}{\bibfnamefont{S.}~\bibnamefont{Xella~Hansen}}
  (\bibinfo{year}{2001}), \bibinfo{note}{in these proceedings}.

\bibitem[{\citenamefont{Battaglia}(2001)}]{Battaglia:2000kd}
\bibinfo{author}{\bibfnamefont{M.}~\bibnamefont{Battaglia}},
  \bibinfo{journal}{Nucl. Instrum. Meth.} \textbf{\bibinfo{volume}{A473}},
  \bibinfo{pages}{75} (\bibinfo{year}{2001}),
  \eprint[http://arXiv.org/abs]{hep-ex/0012021}.

\bibitem[{\citenamefont{Damerell}(2001)}]{Damerell:2001kh}
\bibinfo{author}{\bibfnamefont{C.}~\bibnamefont{Damerell}}
  (\bibinfo{collaboration}{LCFI})  (\bibinfo{year}{2001}),
  \bibinfo{note}{\uppercase{LC-DET-2001-023}}.

\bibitem[{\citenamefont{Burrows}(2001)}]{burrows-ccd}
\bibinfo{author}{\bibfnamefont{P.}~\bibnamefont{Burrows}}
  (\bibinfo{year}{2001}), \bibinfo{note}{in these proceedings}.

\bibitem[{\citenamefont{Brau and Sinev}(2000)}]{Brau:2000pt}
\bibinfo{author}{\bibfnamefont{J.~E.} \bibnamefont{Brau}} \bibnamefont{and}
  \bibinfo{author}{\bibfnamefont{N.}~\bibnamefont{Sinev}},
  \bibinfo{journal}{IEEE Trans. Nucl. Sci.} \textbf{\bibinfo{volume}{47}},
  \bibinfo{pages}{1898} (\bibinfo{year}{2000}).

\bibitem[{\citenamefont{Claus et~al.}(2001)}]{Claus:2001bq}
\bibinfo{author}{\bibfnamefont{G.}~\bibnamefont{Claus}} \bibnamefont{et~al.},
  \bibinfo{journal}{Nucl. Instrum. Meth.} \textbf{\bibinfo{volume}{A473}},
  \bibinfo{pages}{83} (\bibinfo{year}{2001}).

\bibitem[{\citenamefont{Deptuch et~al.}(2001)}]{deptuch}
\bibinfo{author}{\bibfnamefont{G.}~\bibnamefont{Deptuch}} \bibnamefont{et~al.}
  (\bibinfo{year}{2001}), \bibinfo{note}{in these proceedings}.

\bibitem[{\citenamefont{Battaglia
  et~al.}(2001{\natexlab{b}})}]{Battaglia:2001nd}
\bibinfo{author}{\bibfnamefont{M.}~\bibnamefont{Battaglia}}
  \bibnamefont{et~al.}  (\bibinfo{year}{2001}{\natexlab{b}}),
  \eprint[http://arXiv.org/abs]{hep-ex/0102046}.

\bibitem[{\citenamefont{Battaglia and Desch}(2000)}]{Battaglia:2000jb}
\bibinfo{author}{\bibfnamefont{M.}~\bibnamefont{Battaglia}} \bibnamefont{and}
  \bibinfo{author}{\bibfnamefont{K.}~\bibnamefont{Desch}}
  (\bibinfo{year}{2000}), \eprint[http://arXiv.org/abs]{hep-ph/0101165}.

\bibitem[{\citenamefont{Behnke et~al.}(2001{\natexlab{a}})}]{Behnke:2001gb}
\bibinfo{author}{\bibfnamefont{T.}~\bibnamefont{Behnke}} \bibnamefont{et~al.}
  (\bibinfo{year}{2001}{\natexlab{a}}),
  \bibinfo{note}{\uppercase{LC-DET-2001-029}}.

\bibitem[{\citenamefont{Behnke et~al.}(2001{\natexlab{b}})\citenamefont{Behnke,
  Hamann, and Schumacher}}]{Behnke:2001pd}
\bibinfo{author}{\bibfnamefont{T.}~\bibnamefont{Behnke}},
  \bibinfo{author}{\bibfnamefont{M.}~\bibnamefont{Hamann}}, \bibnamefont{and}
  \bibinfo{author}{\bibfnamefont{M.}~\bibnamefont{Schumacher}}
  (\bibinfo{year}{2001}{\natexlab{b}}),
  \bibinfo{note}{\uppercase{LC-DET-2001-006}}.

\bibitem[{\citenamefont{Sauli}(1997)}]{Sauli:1997qp}
\bibinfo{author}{\bibfnamefont{F.}~\bibnamefont{Sauli}},
  \bibinfo{journal}{Nucl. Instrum. Meth.} \textbf{\bibinfo{volume}{A386}},
  \bibinfo{pages}{531} (\bibinfo{year}{1997}).

\bibitem[{\citenamefont{Giomataris et~al.}(1996)\citenamefont{Giomataris,
  Rebourgeard, Robert, and Charpak}}]{Giomataris:1996fq}
\bibinfo{author}{\bibfnamefont{Y.}~\bibnamefont{Giomataris}},
  \bibinfo{author}{\bibfnamefont{P.}~\bibnamefont{Rebourgeard}},
  \bibinfo{author}{\bibfnamefont{J.~P.} \bibnamefont{Robert}},
  \bibnamefont{and} \bibinfo{author}{\bibfnamefont{G.}~\bibnamefont{Charpak}},
  \bibinfo{journal}{Nucl. Instrum. Meth.} \textbf{\bibinfo{volume}{A376}},
  \bibinfo{pages}{29} (\bibinfo{year}{1996}).

\bibitem[{\citenamefont{Gruwe}(1999)}]{Gruwe:1999bd}
\bibinfo{author}{\bibfnamefont{M.}~\bibnamefont{Gruwe}}
  (\bibinfo{year}{1999}), \bibinfo{note}{\uppercase{LC-DET-1999-003-TESLA}}.

\bibitem[{\citenamefont{Ronan and Settles}(2001)}]{tpc-report}
\bibinfo{author}{\bibfnamefont{M.}~\bibnamefont{Ronan}} \bibnamefont{and}
  \bibinfo{author}{\bibfnamefont{R.}~\bibnamefont{Settles}}
  (\bibinfo{year}{2001}), \bibinfo{note}{in these proceedings}.

\bibitem[{\citenamefont{Dima et~al.}(2001)}]{Dima:2001jr}
\bibinfo{author}{\bibfnamefont{M.}~\bibnamefont{Dima}} \bibnamefont{et~al.}
  (\bibinfo{year}{2001}), \bibinfo{note}{in these proceedings},
  \eprint[http://arXiv.org/abs]{hep-ex/0112017}.

\bibitem[{\citenamefont{Brient and Videau}(2001)}]{videau:2001xx}
\bibinfo{author}{\bibfnamefont{J.~C.} \bibnamefont{Brient}} \bibnamefont{and}
  \bibinfo{author}{\bibfnamefont{H.}~\bibnamefont{Videau}}
  (\bibinfo{year}{2001}), \bibinfo{note}{in these proceedings}.

\bibitem[{\citenamefont{Morgunov}(2001)}]{morgunov}
\bibinfo{author}{\bibfnamefont{V.}~\bibnamefont{Morgunov}}
  (\bibinfo{year}{2001}), \bibinfo{note}{in these proceedings}.

\bibitem[{\citenamefont{Brient et~al.}(2001)}]{Brient:2001am}
\bibinfo{author}{\bibfnamefont{J.~C.} \bibnamefont{Brient}}
  \bibnamefont{et~al.}  (\bibinfo{year}{2001}),
  \bibinfo{note}{\uppercase{LC-DET-2001-058}}.

\bibitem[{\citenamefont{Matsunaga}(2001)}]{matsunaga}
\bibinfo{author}{\bibfnamefont{H.}~\bibnamefont{Matsunaga}}
  (\bibinfo{year}{2001}), \bibinfo{note}{in these proceedings}.

\bibitem[{\citenamefont{Soffer}(2001)}]{soffer}
\bibinfo{author}{\bibfnamefont{A.}~\bibnamefont{Soffer}}
  (\bibinfo{year}{2001}), \bibinfo{note}{in these proceedings}.

\bibitem[{\citenamefont{Torrence}(2001)}]{torrence}
\bibinfo{author}{\bibfnamefont{R.}~\bibnamefont{Torrence}}
  (\bibinfo{year}{2001}), \bibinfo{note}{in these proceedings}.

\bibitem[{\citenamefont{Moortgat~Pick}(2001)}]{moortgat:2001xx}
\bibinfo{author}{\bibfnamefont{G.}~\bibnamefont{Moortgat~Pick}}
  (\bibinfo{year}{2001}), \bibinfo{note}{in these proceedings}.

\bibitem[{\citenamefont{Battaglia
  et~al.}(2001{\natexlab{c}})}]{Battaglia:2001zp}
\bibinfo{author}{\bibfnamefont{M.}~\bibnamefont{Battaglia}}
  \bibnamefont{et~al.}, \bibinfo{journal}{Eur. Phys. J.}
  \textbf{\bibinfo{volume}{C22}}, \bibinfo{pages}{535}
  (\bibinfo{year}{2001}{\natexlab{c}}),
  \eprint[http://arXiv.org/abs]{hep-ph/0106204}.

\bibitem[{\citenamefont{Arkani-Hamed et~al.}(1998)\citenamefont{Arkani-Hamed,
  Dimopoulos, and Dvali}}]{Arkani-Hamed:1998rs}
\bibinfo{author}{\bibfnamefont{N.}~\bibnamefont{Arkani-Hamed}},
  \bibinfo{author}{\bibfnamefont{S.}~\bibnamefont{Dimopoulos}},
  \bibnamefont{and} \bibinfo{author}{\bibfnamefont{G.~R.} \bibnamefont{Dvali}},
  \bibinfo{journal}{Phys. Lett.} \textbf{\bibinfo{volume}{B429}},
  \bibinfo{pages}{263} (\bibinfo{year}{1998}),
  \eprint[http://arXiv.org/abs]{hep-ph/9803315}.

\bibitem[{\citenamefont{Antoniadis}(1990)}]{Antoniadis:1990ew}
\bibinfo{author}{\bibfnamefont{I.}~\bibnamefont{Antoniadis}},
  \bibinfo{journal}{Phys. Lett.} \textbf{\bibinfo{volume}{B246}},
  \bibinfo{pages}{377} (\bibinfo{year}{1990}).

\bibitem[{\citenamefont{Randall and Sundrum}(1999)}]{Randall:1999ee}
\bibinfo{author}{\bibfnamefont{L.}~\bibnamefont{Randall}} \bibnamefont{and}
  \bibinfo{author}{\bibfnamefont{R.}~\bibnamefont{Sundrum}},
  \bibinfo{journal}{Phys. Rev. Lett.} \textbf{\bibinfo{volume}{83}},
  \bibinfo{pages}{3370} (\bibinfo{year}{1999}),
  \eprint[http://arXiv.org/abs]{hep-ph/9905221}.

\bibitem[{\citenamefont{Kribs}(2001)}]{Kribs:2001ic}
\bibinfo{author}{\bibfnamefont{G.~D.} \bibnamefont{Kribs}}
  (\bibinfo{year}{2001}), \eprint[http://arXiv.org/abs]{hep-ph/0110242}.

\bibitem[{\citenamefont{Battaglia
  et~al.}(2001{\natexlab{d}})}]{Battaglia:2001xx}
\bibinfo{author}{\bibfnamefont{M.}~\bibnamefont{Battaglia}}
  \bibnamefont{et~al.}  (\bibinfo{year}{2001}{\natexlab{d}}), \bibinfo{note}{in
  these proceedings}, \eprint[http://arXiv.org/abs]{hep-ph/0112270}.

\bibitem[{\citenamefont{Battaglia
  et~al.}(2001{\natexlab{e}})\citenamefont{Battaglia, De~Roeck, and
  Rizzo}}]{Battaglia:2001id}
\bibinfo{author}{\bibfnamefont{M.}~\bibnamefont{Battaglia}},
  \bibinfo{author}{\bibfnamefont{A.}~\bibnamefont{De~Roeck}}, \bibnamefont{and}
  \bibinfo{author}{\bibfnamefont{T.}~\bibnamefont{Rizzo}}
  (\bibinfo{year}{2001}{\natexlab{e}}), \bibinfo{note}{in these proceedings},
  \eprint[http://arXiv.org/abs]{hep-ph/0112169}.

\bibitem[{\citenamefont{Giddings}(2001)}]{Giddings:2001ih}
\bibinfo{author}{\bibfnamefont{S.~B.} \bibnamefont{Giddings}}
  (\bibinfo{year}{2001}), \bibinfo{note}{in these proceedings},
  \eprint[http://arXiv.org/abs]{hep-ph/0110127}.

\bibitem[{\citenamefont{Dimopoulos and Landsberg}(2001)}]{bh}
\bibinfo{author}{\bibfnamefont{S.}~\bibnamefont{Dimopoulos}} \bibnamefont{and}
  \bibinfo{author}{\bibfnamefont{G.}~\bibnamefont{Landsberg}}
  (\bibinfo{year}{2001}), \bibinfo{note}{in these proceedings}.

\bibitem[{\citenamefont{Barklow}(2001)}]{Barklow:2001is}
\bibinfo{author}{\bibfnamefont{T.~L.} \bibnamefont{Barklow}}
  (\bibinfo{year}{2001}), \bibinfo{note}{in these proceedings},
  \eprint[http://arXiv.org/abs]{hep-ph/0112286}.

\bibitem[{\citenamefont{Battaglia
  et~al.}(2001{\natexlab{f}})\citenamefont{Battaglia, Boos, and
  Yao}}]{Battaglia:2001nn}
\bibinfo{author}{\bibfnamefont{M.}~\bibnamefont{Battaglia}},
  \bibinfo{author}{\bibfnamefont{E.}~\bibnamefont{Boos}}, \bibnamefont{and}
  \bibinfo{author}{\bibfnamefont{W.-M.} \bibnamefont{Yao}}
  (\bibinfo{year}{2001}{\natexlab{f}}), \bibinfo{note}{in these proceedings},
  \eprint[http://arXiv.org/abs]{hep-ph/0111276}.

\bibitem[{\citenamefont{Guignard}(2001)}]{clic}
\bibinfo{author}{\bibfnamefont{G.}~\bibnamefont{Guignard}}
  (\bibinfo{year}{2001}), \bibinfo{note}{in these proceedings}.

\bibitem[{\citenamefont{Brinkmass et~al.}(2001)\citenamefont{Brinkmass,
  Raubenheimer, and Toge}}]{m3report}
\bibinfo{author}{\bibfnamefont{R.}~\bibnamefont{Brinkmass}},
  \bibinfo{author}{\bibfnamefont{T.}~\bibnamefont{Raubenheimer}},
  \bibnamefont{and} \bibinfo{author}{\bibfnamefont{N.}~\bibnamefont{Toge}}
  (\bibinfo{year}{2001}), \bibinfo{note}{in these proceedings}.

\end{thebibliography}

\end{document}